\documentclass[aps,pre,twocolumn ,superscriptaddress]{revtex4-1}  % for review and submission  %% superscriptaddress,groupedaddress
\usepackage{graphicx,epsfig}  % needed for figures
\usepackage{dcolumn}   % needed for some tables
\usepackage{bm}        % for math
\usepackage{amssymb}   % for math
\usepackage{color}
\usepackage{mathtools}
\usepackage{float}
\usepackage{makecell}
\usepackage[section]{placeins}

\begin{document}

\title{Enhanced Terahertz Emission from the Wakefield of CO$_2$ Laser-Created Plasma}

%\title{ Upper hybrid resonance heating in laser-plasma interaction in an inhomogenous external magnetic field}

\author{Srimanta Maity}
\email {srimantamaity96@gmail.com}
\affiliation{ELI Beamlines Facility, The Extreme Light Infrastructure ERIC, Za Radnicí 835, 25241 Dolní Břežany, Czech Republic}

\author{Garima Arora}
\email {garimagarora@gmail.com}
\affiliation{Institute of Plasma Physics of the Czech Academy of Sciences, 18200 Prague, Czech Republic}

%~~~~~~~~~~~~~~~~~~~~~~~~~~~~~~~~~~~~~~~~~~~~~~~~~~~~~~~~~~~~~~~~~~~
\begin{abstract}

High-field terahertz (THz) pulse generation is investigated through the interaction of an intense single-color CO$_2$ laser pulse with helium (He) gas targets. Employing multi-dimensional Particle-In-Cell (PIC) simulations, this study reveals a substantial enhancement in THz generation efficiency, even with a single-color laser pulse interacting with gas targets in the self-modulated-laser-wakefield (SMLWF) regime. Our study demonstrates that in the presence of photoionization, a synergistic interplay of laser self-modulation, self-focusing, and local pump depletion leads to the generation of robust THz pulses polarized parallel to the laser electric field. The dependence of THz generation efficiency on target density and laser pulse duration has been investigated. Our study identifies a favourable parametric regime for producing THz fields with amplitudes reaching hundreds of GV/m, surpassing those reported in previous studies.

\end{abstract}

\maketitle

%~~~~~~~~~~~~~~~~~~~~~~~~~~~~~~~~~~~~~~~~~~~~~~~~~~~~~~~~~~~~~~~~~~~
%~~~~~~~~~~~~~~~~~~~~~~~~~~~~~~~~~~~~~~~~~~~~~~~~~~~~~~~~~~~~~~~~~~~
\section{Introduction}\label{intro}
Terahertz (THz) radiation, especially in the frequency gap of 1-10 THz and high field strength, i.e., Gigavolt-per-Meter ( GV/m), has gained wide interest due to many promising applications, e.g., THz spectroscopyv\cite{tonouchi2007cutting,kampfrath2013resonant}, medical imaging \cite{chan2007imaging}, security screening \cite{kemp2003security}, THz-triggered chemistry \cite{larue2015thz}, and biology \cite{alexandrov2013specificity}. The THz spectrum was earlier produced from photonics, such as black body radiation, using electronics and oscillators \cite{ferguson2002materials}. However, these sources only gave nanowatt (nW) power, and their power significantly decreased in this particular gap regime, which even poses more difficulty in getting high power. The high-power THz sources are produced using ultrafast laser-pumped crystals \cite{shalaby2015demonstration,vicario2014generation}. However, they are limited to lower power and bandwidth due to the damaging threshold of the crystal at high power intensity. Quantum cascade lasers \cite{kohler2002terahertz}, photonic crystal lasers \cite{chassagneux2009electrically}, and optically pumped THz gas lasers have also been developed, but they produce tens of mW power. The chase of high peak power has made the technology shift to large-scale accelerators \cite{wu2013intense}; the THz free electron lasers (FEL) \cite{ramian1992new} produce Megawatt (MW) peak power, and the THz from large-scale accelerators can provide GigaWatt (GW) peak powers with GV/m field strength. Nowadays, a new mechanism of THz generation with high power and field strength in the context of intense laser-plasma interaction using a table-top setup has been getting a lot of attention \cite{cook2000intense, clerici2013wavelength, andreeva2016ultrabroad, kim2008coherent}.

A table-top laser-matter interaction is routinely used for electron acceleration \cite{tajima1979laser, joshi2007development, RevModPhys.81.1229, PhysRevApplied.13.034001, PhysRevApplied.18.064091, PhysRevApplied.15.044039, maity2024parametric}, ion acceleration \cite{macchi2013ion, daido2012review}, attosecond pulse generation \cite{PhysRevApplied.16.024042}, higher harmonic generation \cite{ganeev2007high, ganeev2009higher}, inertial confinement fusion \cite{hu2010strong, zylstra2021record, igumenshchev2023proof}, and so on. The THz emission can also be produced in table-top laser-matter interactions when an ultrafast laser pulse interacts with a gas target, even at moderate intensities. Hamster \textit{et al.} \cite{hamster1993subpicosecond} observed the emission at plasma frequency, which is in the THz regime when the fast laser pulse interacts with He gas. The first demonstration of efficient generation of THz pulses was carried out by Cookes \textit{et al.} \cite{cook2000intense} by sending two laser pulses, combining its fundamental frequency with its second harmonic (called two color scheme) in the air. They explained the mechanism of THz generation by four-wave rectification. After that, a few groups reported the THz generation by laser-air interaction using this two-color scheme. The mechanism of THz generation in this scheme is based on the photoionized induced radiation (PIR) proposed by Kim \textit{et al.} \cite{kim2007terahertz} in which the symmetry of positive and negative half-cycles of the fundamental pump wave is broken, generating a net transverse drift current in each ionization instants. Another THz generation approach utilizes laser wakefield excitation in an inhomogeneous plasma. This mechanism is based on the mode conversion from the laser wakefields excited in plasma into electromagnetic waves (THz radiation) \cite{sheng2005emission}. Recently, this scheme has been verified experimentally producing millijoule THz radiation \cite{PhysRevLett.132.165002}. Chen \textit{et. al.} \cite{chen2015high} proposed a new method to generate GV/cm THz field using a single-color ultra-intense laser interaction (i.e., relativistic effects) with near-critical density plasmas. They have shown that due to local pump depletion \cite{decker1996evolution}, the front of an initially Gaussian laser pulse is gradually etched and loses its symmetricity. However, this high field THz suffers very low transmitted efficiency due to absorption from plasma to vacuum boundary. Various studies proposed that an external transverse or longitudinal magnetic field can enhance the transmission efficiency of generated THz emission produced by the intense laser \cite{wang2015tunable,tailliez2022terahertz}.

The abundance of Ti-Saphhire lasers emitting central wavelength 800 nm have been utilized so far for generating THz radiation, and very few studies have been performed for mid and far-infrared wavelengths. Nevertheless, experiments have been conducted that show an increase in THz conversion efficiency using longer wavelengths \cite{koulouklidis2020observation,clerici2013wavelength,nguyen2018broadband}. Wang \textit{et. al.} \cite{wang2011efficient} have theoretically demonstrated that lasers of longer wavelengths can produce stronger THz radiation. Dechard \textit{et al.} \cite{dechard2019thz} studied the wavelength dependence on THz pulse generation by ultraintense two-color laser fields starting from near to far infrared. They have shown that long wavelength (10.6 $\mu$m) significantly enhanced the plasma wakefield, generating tens of GV/m THz radiation in the laser-polarized direction through the photon deceleration mechanism, which fundamentally differs from the well-known PIR mechanism in the two-color scheme for short-wavelength pulses.

Our present study investigates THz generation through laser-matter interactions, employing a system configuration distinct from prior reports. Here, we have shown the THz generation using a single-color CO$_2$ laser pulse interacting with helium gas jets and pre-formed plasmas in the self-modulated laser wakefield (SMLWF) regime \cite{andreev1992resonant, krall1993enhanced, PhysRevLett.72.2887, bulanov1995two, le1997temporal}. The CO$_2$ laser \cite{polyanskiy2015chirped, polyanskiy2020demonstration, panagiotopoulos2020multi}, with a central wavelength in the range 9.6-10.6 $\mu$m, has demonstrated its utility in driving larger wakefields while requiring less power compared to lasers with central wavelengths 0.8-1.0 $\mu$m \cite{kumar2019simulation, brunetti2022high, pogorelsky2016mid, pogorelsky2016bestia}. Additionally, a CO$_2$ laser pulse can also be used to explore the magnetized regime of laser-matter interaction \cite{maity2021harmonic, maity2022mode, juneja2023ion, dhalia2023harmonic, vashistha2023localized}, as the requirement of an external magnetic field in this case is reduced by at least one order of magnitude. This study employs a CO$_2$ laser pulse to bolster THz generation efficiency within a table-top laser-matter interaction setup. Our investigation revealed that a combined effect of photoionization, self-modulation, and local pump depletion drives THz radiation generation. Notably, we have observed THz fields with amplitudes in the hundreds of GV/m range along the laser polarization direction, significantly higher than those in the previously reported studies. In our study, multi-dimensional PIC simulations have been carried out considering a single-color CO$_2$ laser pulse with a central wavelength around 10.6 $\mu$m. We have observed a strong modulation of the laser pulse while propagating through the medium via forward Raman scattering instability \cite{estabrook1980heating, PhysRevLett.47.1285, PhysRevLett.72.1482, fisher1996enhanced, najmudin2000measurement}, generating Stokes and anti-Stokes lines in the optical spectrum. Additionally, electromagnetic radiations of frequencies between 1-10 THz with polarization, the same as the incident laser pulse, have been generated. These THz signals are observed to be transmitted in the vacuum through the right plasma-vacuum boundary along with the transmitted laser pulse. The dependence of THz generation efficiency on the target density and laser pulse duration for long pump wavelength cases, which has not been explored in previous studies, has been investigated. 

This paper is structured as follows. Section \ref{simu} discusses the details of the simulation performed in this study. Section \ref{rd} presents the simulation results and relevant discussion, focusing on generating THz radiation and its underlying mechanisms. Finally, Section \ref{summary} summarizes our research findings, highlighting some conclusive remarks.
%~~~~~~~~~~~~~~~~~~~~~~~~~~~~~~~~~~~~~~~~~~~~~~~~~~~~~~~~~~~~~~~~~~~
%~~~~~~~~~~~~~~~~~~~~~~~~~~~~~~~~~~~~~~~~~~~~~~~~~~~~~~~~~~~~~~~~~~~   
\section{Simulation setup}\label{simu}

%~~~~~~~~~~~~~~~~~~~~~~~~~~~~~~~~~~~~~~~~~~~~~~~~~~~~~~~~~~~~~~~~~~~ 

\begin{figure}[hbt!]
  \centering
  \includegraphics[width=3.2in]{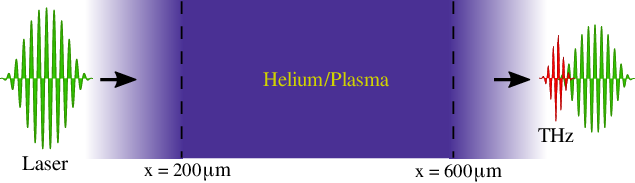}
  \caption{A schematic of the simulation setup has been shown. A CO$_2$ laser pulse with 10.6 $\mu$m central wavelength is incident from the simulation box's left-boundary ($x = 0$) on a helium (He) gas target. The He gas target has a trapezoidal density profile along the laser propagation direction, i.e., $\hat x$ with a 400-$\mu$m-long plateau (in between the dashed vertical lines at $x = 200$ $\mu$m and 600 $\mu$m) and 25 $\mu$m ramps. After the interaction, a terahertz (THz) signal is observed to be transmitted through the right plasma-vacuum boundary, propagating in the forward direction along with the transmitted laser pulse.}
\label{fig_schmtc}
\end{figure}
%~~~~~~~~~~~~~~~~~~~~~~~~~~~~~~~~~~~~~~~~~~~~~~~~~~~~~~~~~~~~~~~~~~~

\begin{table*}
\caption{Simulation parameters}
\vspace{0.25cm}
\label{table1}
\resizebox{\textwidth}{!}{%
\begin{tabular}{|c|c|c|c|}
\hline
    \makecell{Laser wavelength\\ ($\lambda_0$)} & \makecell{Laser spot size (FWHM)\\ ($w_{fwhm}$)} & \makecell{Laser pulse duration (FWHM)\\ ($\tau_{fwhm}$)} & \makecell{Peak value of normalized\\ vector potential ($a_0$)}\\ 
\hline
    10.6 $\mu$m & 50 $\mu$m & 100-500 fs & 5.0\\ 
\hline
\hline
\hline
    \makecell{Laser frequency ($\nu_0$)} &  \makecell{Laser peak power ($P_0$)} & Laser peak intensity ($I_0$) & \makecell{Laser peak energy ($E_0$)}\\ 
\hline
    $28.3$ THz & $4.3$ TW & $3.0\times 10^{17}$ W/cm$^2$ & $\sim$ 0.4-2.1 J\\     
\hline
\hline
\hline
\makecell{Helium density in the plateau \\ ($n_{He}$)} & \makecell{Critical density of \\ plasma electrons ($n_c$)} & \makecell{Plasma frequency ($\nu_{pe}$) \\ corresponding to $n_{He}$} & $\nu_{0}/\nu_{pe}$\\ 
\hline
    $(0.05-0.5)\times 10^{18}$ cm$^{-3}$ & $\sim$ $1.0\times 10^{19}$ cm$^{-3}$ & $\sim$ 3-9 THz & $\sim$ 10-3\\
\hline        
\end{tabular}}
\end{table*}
%~~~~~~~~~~~~~~~~~~~~~~~~~~~~~~~~~~~~~~~~~~~~~~~~~~~~~~~~~~~~~~~~~~~
A set of 2D3V (2D in space and 3D in velocity) PIC simulations has been performed using a fully relativistic, massively parallelized, open source PIC code, EPOCH \cite{arber2015contemporary, bennett2017users}. Second-order Yee scheme \cite{yee1966numerical} and Boris rotation algorithm \cite{boris1970relativistic} with a modified leapfrog method are used for \textit{field solver} and  \textit{particle pusher}, respectively. A 2D rectangular Cartesian geometry (x-y) has been considered as a simulation domain. The simulation box is extended from $0$ to $100\lambda_0$ along $\hat x$ and from $-50\lambda_0$ to $50\lambda_0$ along $\hat y$. Here, $\lambda_0$ represents the laser wavelength. The grid size is considered to be $\lambda_0/30$ $\times$ $\lambda_0/30$, corresponding to the $3000$ number of grid cells in each direction. The simulation time step is considered $\Delta t = 0.58$ fs, which is low enough to resolve the time scales of all the physical processes of our interest in this study. To reduce numerical dispersion, Courant number  ($C$), following the CFL (Courant-Friedrichs-Lewy) condition ($C =  c\Delta t /\Delta x + c\Delta t /\Delta y \leq 1$) was chosen to be $C = 0.99$, i.e., close to 1.0. Here $c$ defines the speed of light in vacuum, and $\Delta x$, $\Delta y$ represent the grid size along $\hat x$, $\hat y$ direction, respectively. In our simulations, sixteen simulation particles are initially considered in each cell. Open boundary conditions are used in all four directions for both the electromagnetic fields and particles.

The simulation configuration considered in this study has been illustrated by the schematic in Fig. \ref{fig_schmtc}. Helium gas (He), or a preformed plasma channel, is considered as a target. The target has a trapezoidal density profile along $\hat x$ with a 400-$\mu$m-long plateau and $25$ $\mu$m ramps. The plateau starts from $x = 200$ $\mu$m and extends up to $x = 600$ $\mu$m. In EPOCH \cite{arber2015contemporary}, different ionization models are included to account for various modes of the ionization process, e.g., collisional ionization and field ionization. We have not considered collisions between the simulation particles, as their effect can be ignored for higher irradiances ($I_0\lambda_0^2\geq 10^{15}$ Wcm$^{-2}$$\mu$m$^{2}$) of the laser pulse \cite{gibbon2005short}. However, in our simulations, we have incorporated field ionization modules, i.e., multi-photon ionization \cite{delone2000multiphoton} and tunneling ionization based on the Ammosov-Delone-Krainov (ADK) formula \cite{ammosov1986tunnel} as well as barrier-suppression ionization (BSI) process \cite{krainov1995theory}.

A single-color Gaussian laser pulse with central wavelength $\lambda_0 = 10.6$ $\mu$m is considered. The laser starts from the simulation box's left boundary ($x = 0$) and propagates along $+\hat x$ direction. The laser is polarized along $\hat z$ and initially focused at $x = 200$ $\mu$m, i.e., at the beginning of the target plateau. We have considered a Gaussian profile of the laser pulse in the transverse plane (along $\hat y$) with a Full-Width-Half-Maximum (FWHM) spot size $w_{fwhm} = 50$ $\mu$m at focus. In our simulations, the laser pulse duration has been varied from $\tau_{fwhm} = 100$ to 500 fs. The initial peak value of normalized vector potential is considered to be $a_0 = 5.0$ at focus, which corresponds to the laser peak intensity $I_0 \approx 3.0\times 10^{17}$ W/cm$^2$ and peak power $P_0 \approx 4.3$ TW. The central frequency of the laser pulse is $\nu_0 = \omega_0/2\pi \approx 28$ THz. Thus, the value of critical density of plasma electrons, above which the laser pulse can no longer penetrate (beyond the skin depth distance) the target, becomes $n_c \approx 1.0\times 10^{19}$ cm$^{-3}$. Table \ref{table1} also provides all the relevant simulation parameters.

%~~~~~~~~~~~~~~~~~~~~~~~~~~~~~~~~~~~~~~~~~~~~~~~~~~~~~~~~~~~~~~~~~~~
%~~~~~~~~~~~~~~~~~~~~~~~~~~~~~~~~~~~~~~~~~~~~~~~~~~~~~~~~~~~~~~~~~~~
\section{Results and Discussion}
\label{rd}

%~~~~~~~~~~~~~~~~~~~~~~~~~~~~~~~~~~~~~~~~~~~~~~~~~~~~~~~~~~~~~~~~~~~

In the present study, we have investigated the interaction of a CO$_2$ laser pulse with a helium (He) gas target in the self-modulated laser wakefield (SMLWF) regime. As the laser pulse propagates through the neutral He target, it ionizes the medium and forms a plasma channel. As soon as electrons are ionized, they start to experience laser ponderomotive force ($F_p \propto \nabla{I}/m$) \cite{kruer1988physics} and are eventually expelled from the region of the laser pulse. The generated ions also experience the laser ponderomotive force in the same direction as the electrons. However, ions remain stationary in the electron timescale because of their higher inertia. As a result, large amplitude plasma waves (wakefields) are excited with a phase velocity approximately equal to the group velocity of the laser pulse inside the medium. The parametric regime where the longitudinal length of the laser pulse $c\tau_{fwhm}$ becomes longer than the plasma wavelength $\lambda_p$ is known as the self-modulated laser wakefield (SMLWF) regime. In the SMLWF regime, even for the small-amplitude density perturbation originating from the ponderomotive force,  the laser pulse encounters a periodically fluctuating index of refraction within the medium, rather than a constant one. As a result, the laser pulse envelope gets modulated at plasma wavelength $\lambda_p$ generating beats. The ponderomotive force associated with this modulated laser pulse resonantly amplifies the plasma wave, leading to a higher density perturbation, thereby producing an instability known as self-modulation instability \cite{andreev1992resonant, krall1993enhanced, PhysRevLett.72.2887}. As a signature of this instability, the transmitted laser spectrum exhibits forward Raman scattered sidebands at ($\omega_0 + \omega_{pe}$, $k_0 + k_{p}$ (anti-Stokes line) and ($\omega_0 - \omega_{pe}$, $k_0 - k_{p}$) (Stokes line). Here, $\omega_0 = 2\pi \nu_0$ and $k_0 = 2\pi/\lambda_0$ denote the fundamental frequency and wavenumber of the laser pulse, while $\omega_{pe} = 2\pi\nu_{pe}$ and $k_{p} = \omega_{pe}/c$ represent the plasma frequency and its associated wavenumber, respectively. Our present simulation study shows that in addition to the Raman scattered sidebands, electromagnetic (EM) signals in the frequency range $1-10$ THz are also generated in the interaction process. Below, we have discussed the mechanism and parametric dependence of these generated THz radiation.

%~~~~~~~~~~~~~~~~~~~~~~~~~~~~~~~~~~~~~~~~~~~~~~~~~~~~~~~~~~~~~~~~~~~

\begin{figure}[hbt!]
  \centering
  \includegraphics[width=3.3in]{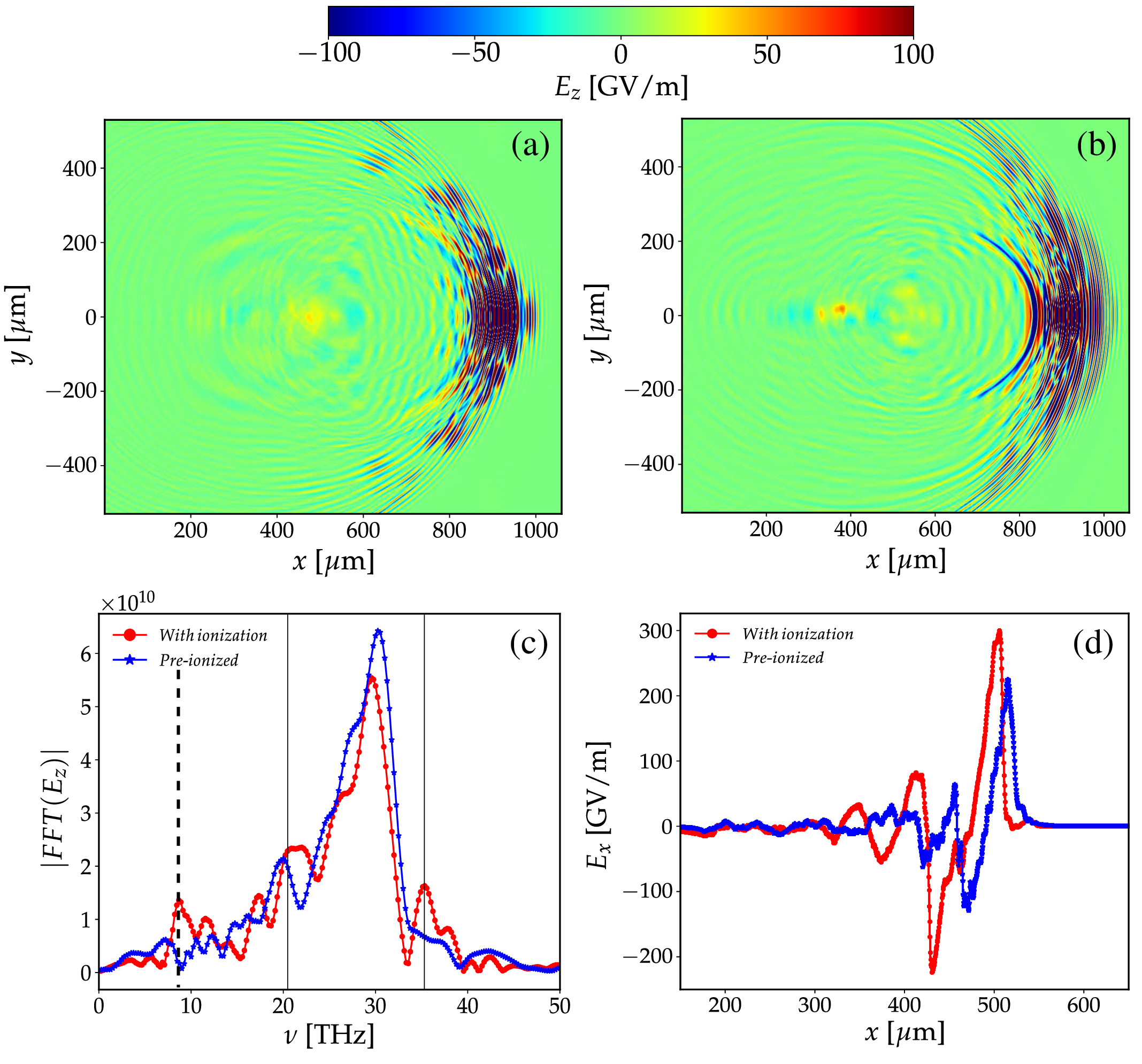}
  \caption{ (a)-(b) Distribution of electric field $E_z$ in the x-y plane produced by a single-color CO$_2$ laser pulse with $\lambda_0 = 10.6$ $\mu$m and $a_0 = 5.0$ for (a) ionizing He (i.e., with photoionization) and (b) pre-ionized plasma target at $t = 3.5$ ps. (c) Fourier spectra of $E_z(t)$ evaluated at $x = 700$ $\mu$m (i.e., $\sim$100 $\mu$m after the target) for ionizing He (red circle-marked line) and pre-ionized plasma (blue star-marked line). Here, two solid black vertical lines locate the peak positions of anti-Stokes and Stokes sidebands, and a dashed vertical line identifies the THz peak. (d) Longitudinal electric field $E_x$ inside the target for these two cases at $t = 2.0$ ps. In both cases, the plateau electron density is $n = 0.10n_c$.}
\label{fig_preplsma}
\end{figure}

%~~~~~~~~~~~~~~~~~~~~~~~~~~~~~~~~~~~~~~~~~~~~~~~~~~~~~~~~~~~~~~~~~~~

First, we investigate the impact of photoionization in the interaction process in the SMLWF regime by comparing results from two scenarios. In the first scenario, a CO$_2$ laser propagates through a neutral helium gas target. In the second case, we consider the laser interaction with a pre-ionized plasma having an electron density profile ($n = 0.10n_c$) identical to that generated from the photoionizing helium. The simulation results in these two cases are shown in Fig. \ref{fig_preplsma}. The distribution of laser electric field $E_z$ transmitted to the vacuum after propagating through the target is demonstrated in Fig. \ref{fig_preplsma}(a), (b) for pre-ionized plasma and neutral He targets, respectively. The 2D profiles of transmitted $E_z$ exhibit distinct differences between these two cases, implying a modification in the interaction process due to photoionization. To dig into more details, we have performed Fourier transformation of the on-axis transmitted laser field ($E_z$) and presented in Fig. \ref{fig_preplsma}(c). In both the cases, i.e., pre-ionized plasma and neutral He targets, in addition to the fundamental frequency at $\nu_0 \sim 28$ THz, a strong Stokes sideband at $\nu = \nu_0 -\nu_{pe}$ ($\sim 20$ THz) appears in the corresponding Fourier spectrum. However, the anti-Stokes sideband ($\nu = \nu_0 + \nu_{pe}$) is more apparent in the case of photoionizing He. A similar observation was also reported by Kumar \textit{et al.} \cite{kumar2019simulation}. Moreover, our investigation unveils a notable feature in the Fourier spectrum, showcasing a prominent peak at a frequency $\nu<10$ THz (indicated by a dashed vertical line). Importantly, this peak exhibits higher intensity in the presence of photoionizing He than the pre-ionized target. Notably, this observation was not addressed in the previous study by Kumar \textit{et al.} \cite{kumar2019simulation}, which considered a hydrogen target. The detailed mechanism and analysis of the THz emission will be explored in subsequent discussions. However, an understanding of the effect of photoionization in the interaction process can be elucidated as follows. In the presence of an ionizing front, the dielectric function of a medium undergoes temporal variation rather than remaining constant. Mori \textit{et al.} \cite{mori1992ponderomotive} conducted a theoretical investigation into the laser-driven ponderomotive force within such a medium. Their findings suggest that when there is a co-moving ionization front, the ponderomotive force amplifies significantly compared to a pre-ionized medium. As a result, the wakefield ($E_x$) amplitude becomes higher in the case of photoionizing He compared to the pre-ionized He plasma target, as illustrated in Fig. \ref{fig_preplsma}(d). The stronger wakefield amplifies the initiation of the forward Raman scattering instability, enhancing the intensity of the Stokes and anti-Stokes sidebands. 

Thus, it has been established that photoionization plays a distinct and significant role in the interaction process. Moreover, we will demonstrate the pivotal contribution of photoionization to generating THz radiation. From now on, our analysis will focus primarily on THz radiation and its parametric dependencies.

%~~~~~~~~~~~~~~~~~~~~~~~~~~~~~~~~~~~~~~~~~~~~~~~~~~~~~~~~~~~~~~~~~~~

\begin{figure}[hbt!]
  \centering
  \includegraphics[width=3.0in]{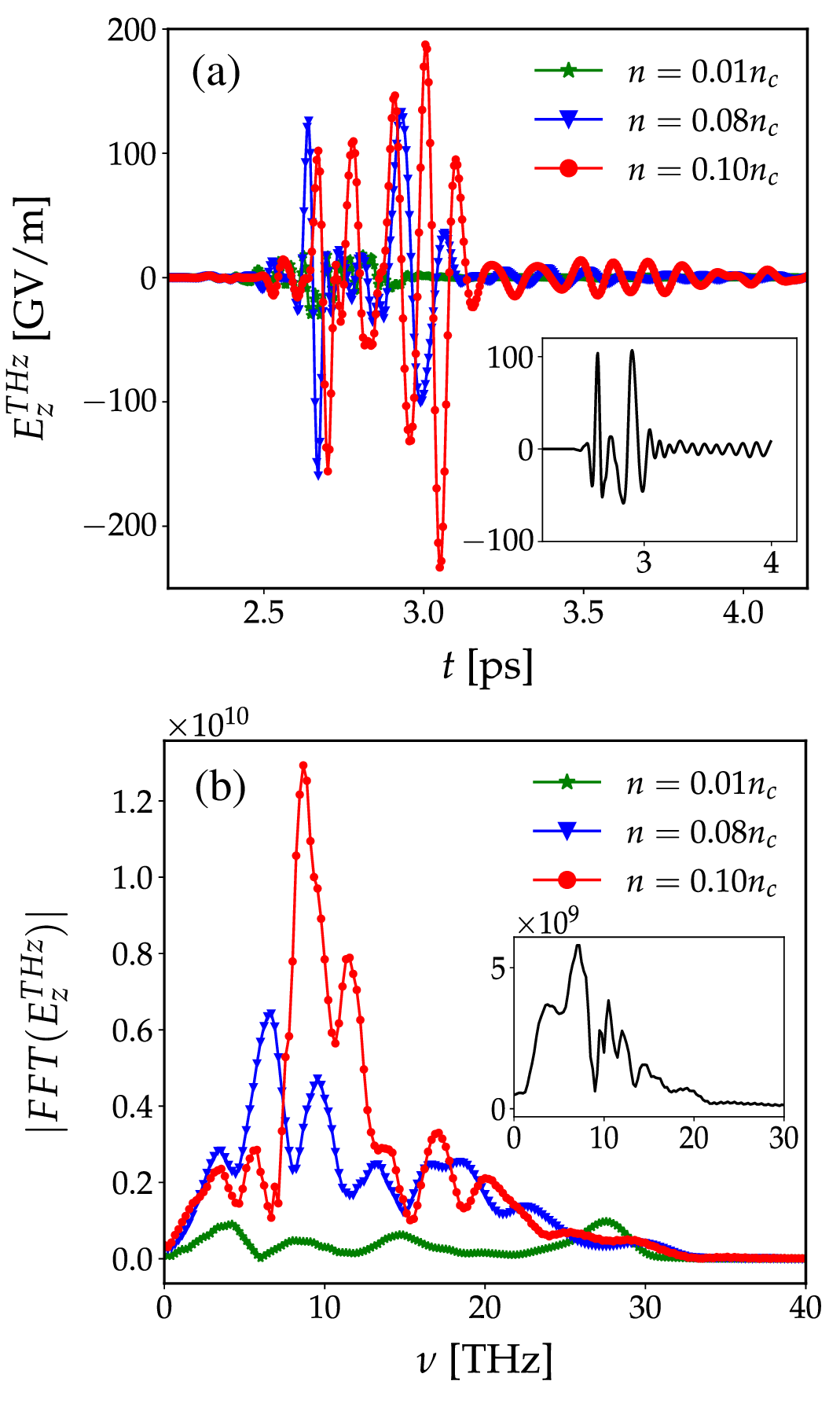}
  \caption{ (a) Terahertz signal ($E_z^{THz}$) extracted from the on-axis ($y = 0$) $E_z(t)$ transmitted to vacuum at $x = 700$ $\mu$m after filtering in the frequency range $\nu<\nu_0/2$ for simulations considering ionizing He with different values of electron density ($n$) at the plateau. (b) Corresponding Fourier spectra of $E_z^{THz} (t)$ for these cases. Insets in (a) and (b) show the extracted (at $x = 700$ $\mu$m) on-axis THz signal and its Fourier spectrum for pre-ionized plasma target with plateau electron density $n = 0.1n_c$, respectively. We have considered the laser pulse duration $\tau_{fwhm} = 300$ fs for all the cases.}
\label{fig_thz_n0}
\end{figure}

%~~~~~~~~~~~~~~~~~~~~~~~~~~~~~~~~~~~~~~~~~~~~~~~~~~~~~~~~~~~~~~~~~~~

The THz signals transmitted into the vacuum ($x=700$ $\mu$m) have been shown in Fig. \ref{fig_thz_n0}(a) for different simulations with changing values of plateau (electron) density $n$. Fig. \ref{fig_thz_n0}(b) illustrates the corresponding Fourier spectra. In the insets of Fig. \ref{fig_thz_n0}(a) and (b), THz signals and corresponding Fourier spectrum have been shown for the case with pre-ionized plasma with electron density at the plateau $n = 0.10n_c$. The THz signals have been extracted from the transmitted on-axis $E_z$ (t) at the location $x = 700$ $\mu$m performing inverse Fourier Transform within the frequency window $\nu\equiv\omega/2\pi<\nu_0/2$. It is seen that THz fields with 200 GV/m maximum amplitude transmit to the vacuum along with the remaining laser field after the interaction. The amplitude of these THz fields increases with an increase in target density. For $n \lesssim 0.01n_c$, the THz field strength becomes negligible. It is also seen (from the inset of Fig. \ref{fig_thz_n0}) that for the same plateau electron density, THz field strength further enhances when a neutral medium (e.g., He) is considered instead of a pre-ionized plasma. The Fourier spectra of these THz radiations illustrated in Fig. \ref{fig_thz_n0}(b) show peaks around the corresponding relativistic plasma frequency $\Tilde{\nu}_{pe} \equiv \nu_{pe}/\sqrt{\gamma} < 10$ THz, where $\gamma$ defines the Lorentz factor.

%~~~~~~~~~~~~~~~~~~~~~~~~~~~~~~~~~~~~~~~~~~~~~~~~~~~~~~~~~~~~~~~~~~~

\begin{figure*}[hbt!]
  \centering
  \includegraphics[width=6.4in]{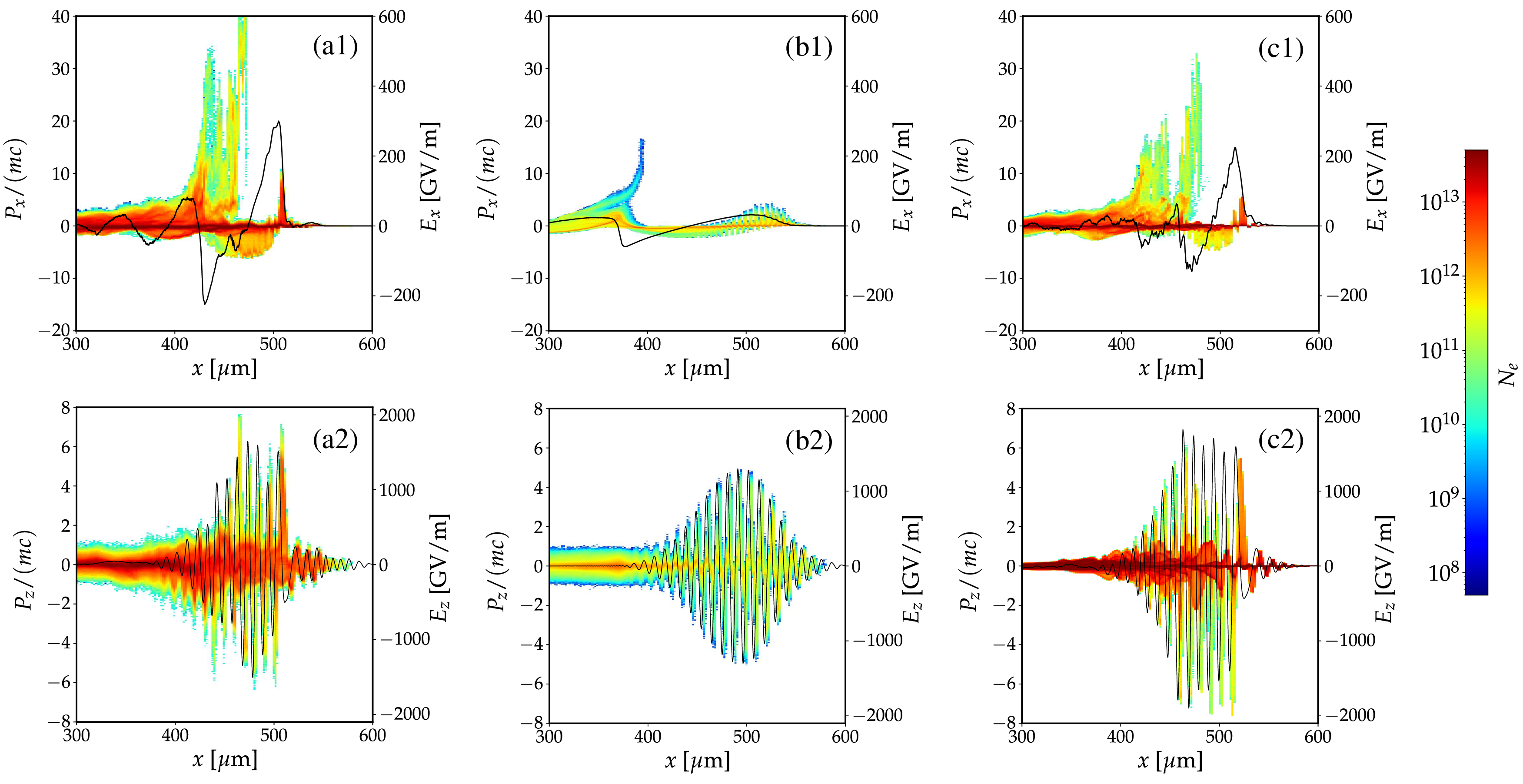}
  \caption{ (a1), (b1), and (c1) Longitudinal momentum $p_X$ (color patches) with on-axis longitudinal electric field $E_x$ (black solid curves) at $t = 2.0$ ps. (a2), (b2), and (c2) Transverse momentum $p_z$ (color patches) along with on-axis electric field $E_z$ (black solid curves) at $t = 2.0$ ps. For (a1)-(a2) and (b1)-(b2), ionizing He was considered with $n = 0.1n_c$ and $0.01n_c$, respectively. Whereas, for (c1)-(c2), pre-formed plasma channel with $n = 0.1n_c$ was considered. Here, the label $N_e$ in the colorbar represents the number of ionized electrons. We have considered $\tau_{fwhm} = 300$ fs for all the cases.}
\label{fig_px_n0}
\end{figure*}

%~~~~~~~~~~~~~~~~~~~~~~~~~~~~~~~~~~~~~~~~~~~~~~~~~~~~~~~~~~~~~~~~~~~

To comprehensively understand our observations, we delve into the target during its interaction with the laser pulse. Specifically, we have examined the phase-space distribution of electrons as well as longitudinal and transverse electric fields. Figures \ref{fig_px_n0}(a1)-\ref{fig_px_n0}(b1) depict a comparison between the longitudinal phase-space ($P_x$-$x$) distributions of electrons and the on-axis longitudinal electric field $E_x$ at a particular time ($t = 2.0$ ps) for two different plateau densities $n = 0.1n_c$ and $0.01n_c$, respectively, considering photoionization of helium in both scenarios. In both cases, the laser pulse duration is considered to be $\tau_{fwhm} = 300$ fs. It is seen that wakefield amplitude and, consequently, the longitudinal momentum gained by electrons are much higher for higher target density since $E_x\propto \sqrt{n}$. Electrons are also observed being injected into the accelerating phase of the wakefield for $n=0.1n_c$, which will be accelerated further as the laser propagates through the medium. These accelerated electrons will radiate x-ray through betatron motion inside the wakefield \cite{PhysRevLett.118.134801}. Upon crossing the plasma-vacuum interface, they will also emit radially polarized coherent transition radiation (CTR) in the THz band range \cite{PhysRevLett.91.074802, dechard2019thz}. However, our present study mainly focuses on the THz radiation in the laser polarization direction (i.e., $\hat z$).

Transverse momentum ($P_z$) of electrons (color patches) and on-axis laser field $E_z$ (black solid curves) have been depicted as a function of x ($t = 2.0$ ps) in Fig. \ref{fig_px_n0}(a2) and Fig. \ref{fig_px_n0}(b2) for these two cases, i.e., $n = 0.1n_c$ and $0.01n_c$, respectively. Although the laser intensity ($a_0$) was the same for both cases, the peak values of $E_z$ are significantly larger for higher plateau density ($n=0.10n_c$). This outcome is attributed to the combined effect of relativistic self-focusing \cite{sprangle1987relativistic, PhysRevA.41.4463} and ponderomotive self-channeling \cite{sun1987self, PhysRevA.40.3230} of laser pulse inside the target. The relativistic self-focusing occurs when laser power exceeds a critical power ($P_c$), i.e., $P_0\gtrsim P_c = 17\omega_0^2/\omega_{pe}^2$ GW. It is based on the change of plasma refractive index through the relativistic mass correction of electrons. However, this mechanism is mainly effective for a long laser pulse, i.e.,  $\tau_{fwhm}>\omega_{pe}^{-1}$, as the modification of the index of refraction by the laser pulse requires a time $\sim \omega_{pe}^{-1}$. Additionally, for a long laser pulse ($L = c\tau_{fwhm}>\lambda_p$), ponderomotive self-channeling enhances the relativistic self-focusing effect. This is because, as the laser propagates through the plasma, it pushes away electrons from its path, creating a density channel. Consequently, the radial gradient (from the laser propagation axis) of the refractive index becomes negative, leading to the self-focusing of the laser spot size. Even though both scenarios with $n = 0.1n_c$ and $n = 0.01n_c$ satisfy the condition $P_0>P_c$, it is noteworthy that only in the former case does $L>\lambda_p$. Consequently, the on-axis laser electric field $E_z$ and thus $P_z$, demonstrate significantly higher values for $n = 0.1n_c$, as evident in Fig. \ref{fig_px_n0}(a2)-(b2). Additionally, for the case with $n = 0.1n_c$, as the longitudinal length of the laser pulse ($c\tau_{fwhm}$) is longer compared to the plasma wavelength ($\lambda_p$), the laser pulse gets modulated via self-modulation instability \cite{andreev1992resonant}, as can be seen from Fig. \ref{fig_px_n0}(a2). In the self-modulated regime, the constant feedback of wake on the pulse, coupled with the self-consistent evolution of both, results in the self-steepening and self-shortening of the laser pulse. These combined effects enhance the wakefield amplitude for higher plateau density. It is to be noticed that all these phenomena occur even in a preformed plasma medium with higher density (e.g., $n = 0.1n_c$), as depicted in Fig. \ref{fig_px_n0}(c1)-(c2). An additional contribution from photoionization is that the wakefield amplitude and, thus, longitudinal momentum ($P_x$) gained by the electrons increases \cite{dechard2019thz}, as discussed previously and also can be seen from Fig. \ref{fig_px_n0}(a1) $\&$ (c1). Furthermore, at higher densities, with and without photoionization, the leading edge of the laser envelope etches back and erodes due to the effect of local pump depletion. Moreover, in higher density targets where $L>\lambda_p$, the laser pulse undergoes self-modulation instability, resulting in the breakdown of symmetry between the positive and negative cycles of the pulse. These phenomena are illustrated in Fig. \ref{fig_px_n0}(a2) and (c2). Conversely, these effects are absent in lower-density targets, allowing the laser envelope to maintain its Gaussian symmetry, as depicted in Fig. \ref{fig_px_n0}(b2).

%~~~~~~~~~~~~~~~~~~~~~~~~~~~~~~~~~~~~~~~~~~~~~~~~~~~~~~~~~~~~~~~~~~~

\begin{figure}[hbt!]
  \includegraphics[width=3.1in]{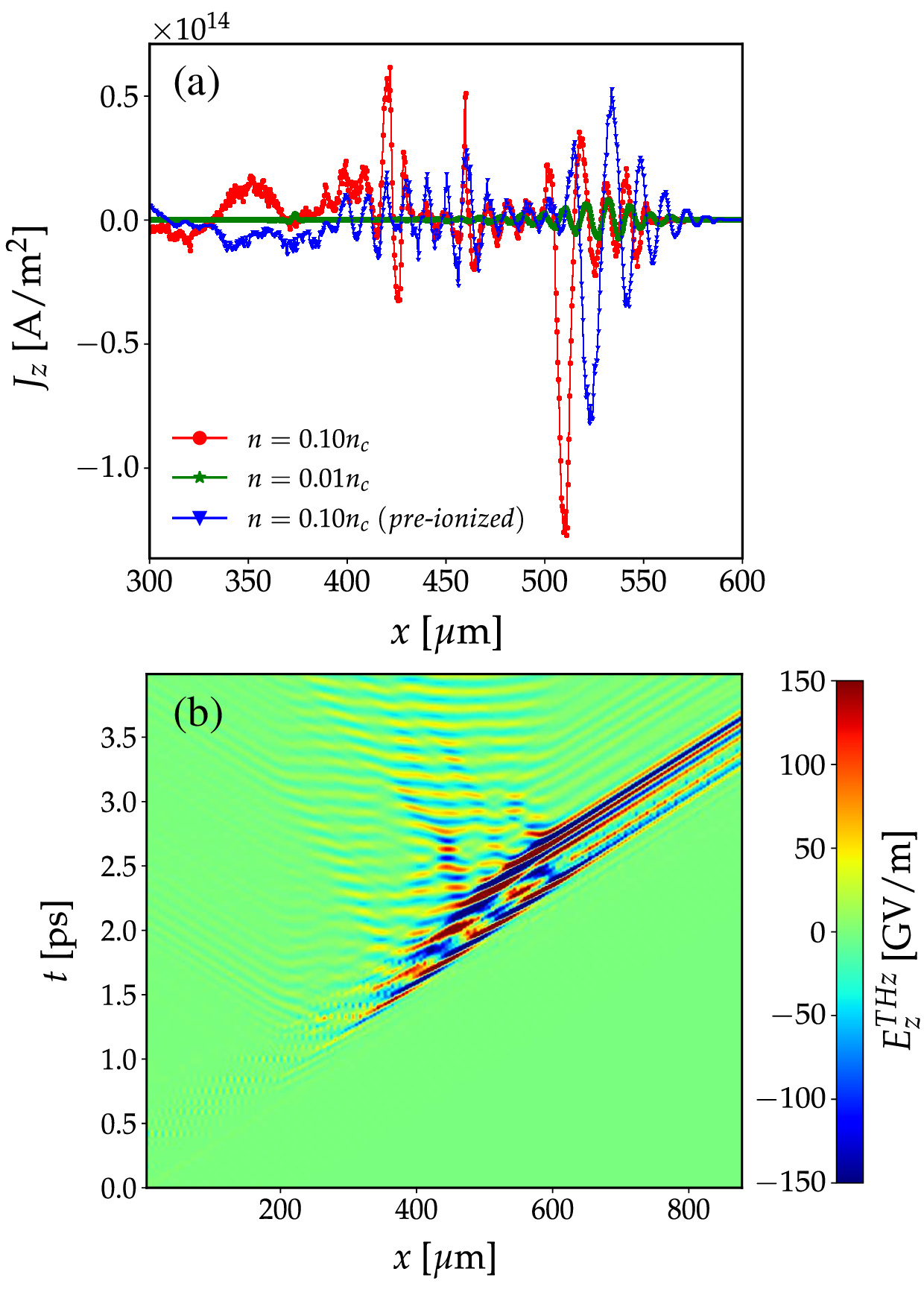}
  \caption{ (a) On-axis transverse current ($J_z$) profile along the laser propagation direction ($\hat x$) at $t = 2.0$ ps for different values of target density with and without photoionization. (b) Distribution of THz field in t-x plane extracted from $E_z(t)$ in the frequency range $\nu<\nu_0/2$ for $n = 0.10n_c$.}
\label{fig_j_pclr_thz}
\end{figure}

%~~~~~~~~~~~~~~~~~~~~~~~~~~~~~~~~~~~~~~~~~~~~~~~~~~~~~~~~~~~~~~~~~~~

The THz radiation is generated through the transverse drift current modulated at relativistic plasma frequency. In our present study, a combined effect of different physical processes produces this transverse drift current. The transverse current ($J_z$) as a function of $x$ has been shown in Fig. \ref{fig_j_pclr_thz}(a) at a particular time $t = 2.0$ ps. In the case of low-density targets (illustrated by the green star-marked curve), the generated transverse current is negligible, confined mainly in the ionization zone, approximately around $x \sim 500$ $\mu$m. Conversely, with higher density, whether with or without photoionization, the transverse current $J_z$ not only appears within the ionization zone but also extends to the rear pulse ($x<400$ $\mu$m). Notably, photoionization (red circle-marked curve) enhances the $J_z$. In the case of preformed plasma where photoionization is absent, the net transverse drift current is produced due to the breaking of the symmetry of the Gaussian laser envelope through the self-modulation and local pump depletion effect \cite{chen2015high, decker1996evolution}, as demonstrated in Fig. \ref{fig_px_n0}(c2). In the case of neutral helium, besides the effects mentioned above,  photoionization significantly contributes to the generation of transverse drift currents. For a low-intense, short-wavelength laser pulse, ionization occurs over many laser cycles around the peak of the envelope. Consequently, the net transverse momentum generated by ionization is effectively averaged out. However, when the intensity of the laser is increased, and its wavelength is longer, photoionization can occur at the front edge within a single cycle or even half a cycle of the laser, resulting in a net transverse momentum of electrons associated with each ionization event \cite{wang2011efficient}. At higher target density, the dynamic interplay between laser self-focusing, self-modulation instability, and local pump depletion synergistically amplifies the generation mechanism of transverse drift current associated with photoionization. The combined effect results in heightened laser intensity, resonant amplification of density accumulation, and erosion of the laser front edge. Consequently, it substantially reinforces the underlying process.

The temporal and spatial evolution of THz radiation generation for a particular case considering photoionizing He target with $n = 0.1n_c$ has been traced and illustrated in Fig. \ref{fig_j_pclr_thz}(b). The laser pulse duration is fixed at $\tau_{fwhm} = 300$ fs. As mentioned before, here also, THz fields are extracted by employing a low-pass filter on the on-axis $E_z(x,t)$ within the frequency range $\nu<\nu_0/2$. It is observed that THz radiation generation initiates after the laser has propagated through a certain distance ($x \approx 400$ $\mu$m) within the target. The intensity of the resulting THz fields demonstrates a steady growth as the laser pulse advances through the plateau region of the target (up to $x = 600$ $\mu$m). These observations are consistent with our previously outlined mechanism for THz generation. The effect of self-focusing, self-modulation instability, and local pump depletion manifests notably only after a certain distance of propagation. It becomes more and more prominent as the laser continues its propagation within the target. Despite being considerably higher than previous reports, only a fraction of the generated THz fields transmit through the plasma-vacuum interface. The transmission efficiency can be controlled by applying an external magnetic field and adjusting the exit ramp \cite{tailliez2022terahertz}. Additionally, as depicted in Fig. \ref{fig_j_pclr_thz}(b), the generated THz radiation predominantly propagates forward alongside the laser pulse. This observation is consistent with the previously reported studies.
%~~~~~~~~~~~~~~~~~~~~~~~~~~~~~~~~~~~~~~~~~~~~~~~~~~~~~~~~~~~~~~~~~~~

\begin{figure}[hbt!]
  \centering
  \includegraphics[width=3.4in]{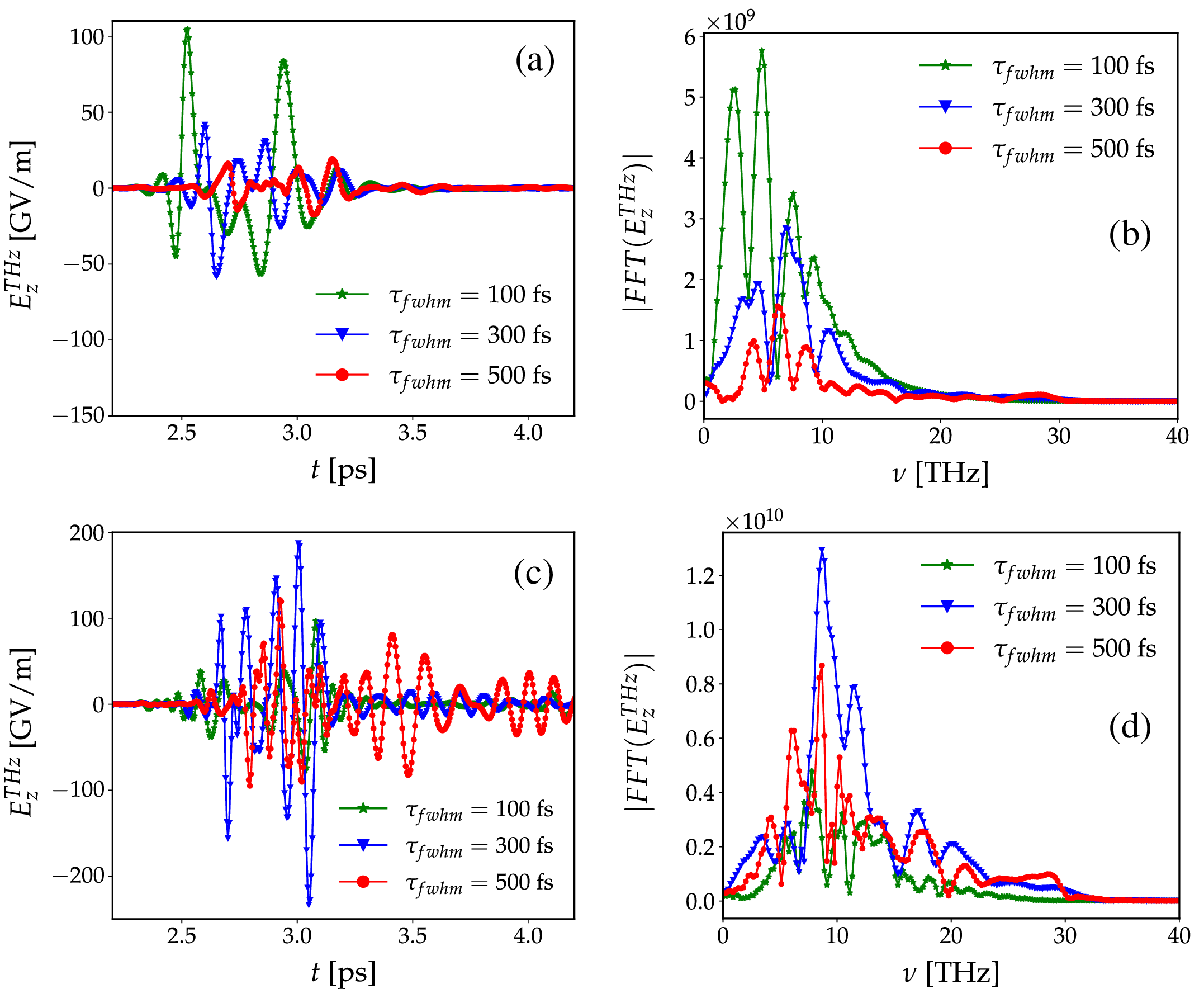}
  \caption{ (a), (b) Extracted THz signal $(E_z^{THz})$ from on-axis electric field $E_z$ transmitted to the vacuum (at $x = 700$ $\mu$m) and corresponding Fourier spectra for three different values of laser pulse duration ($\tau_{fwhm}$) with a fixed target (photoionized He electron) plateau density $n = 0.05n_c$, respectively. (c), (d) Same information for $n = 0.1n_c$.}
\label{fig_thz_tau}
\end{figure}

\begin{figure}[hbt!]
  \centering
  \includegraphics[width=3.3in]{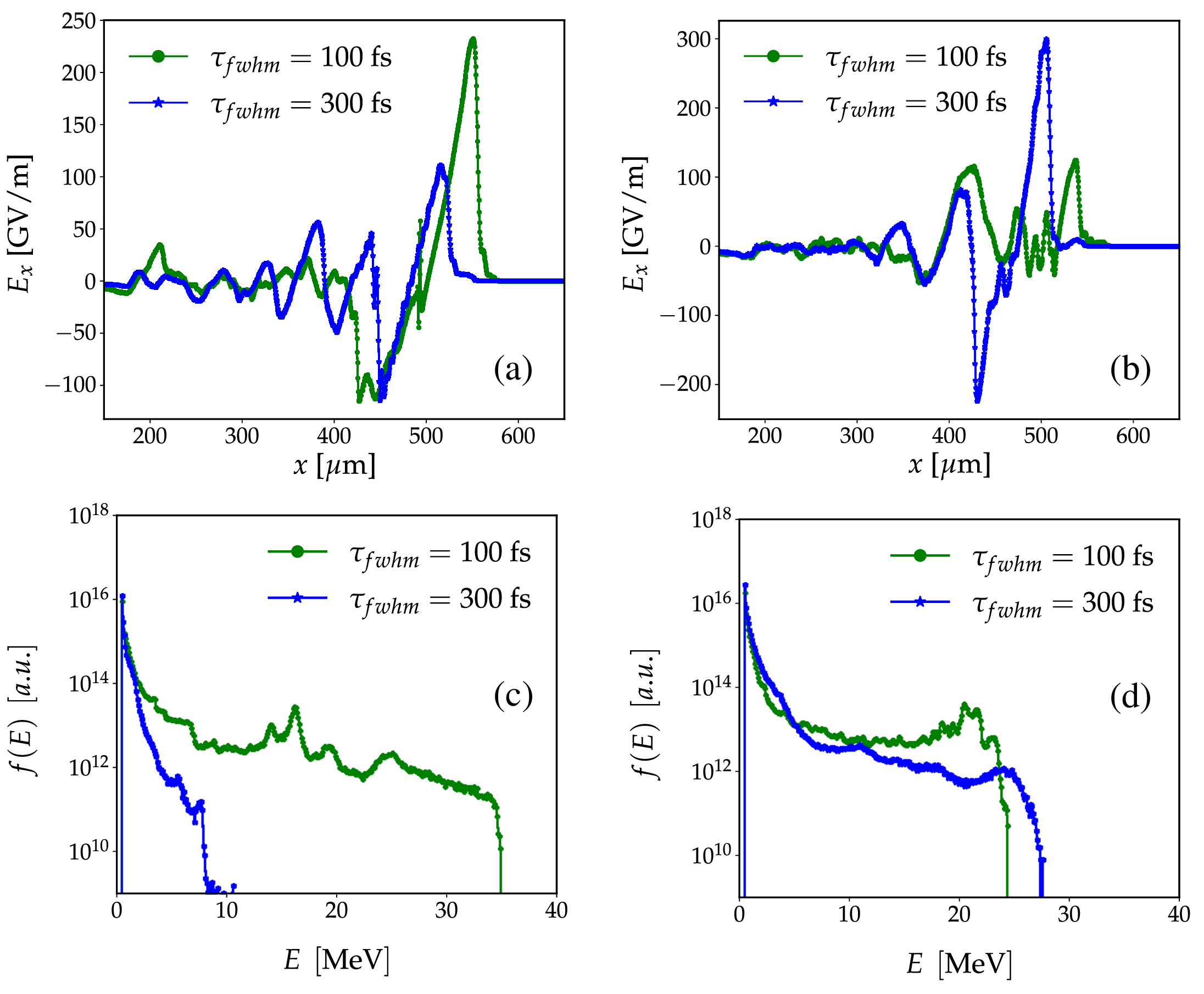}
  \caption{ (a), (b) On-axis longitudinal electric field $E_x$ at $t = 2.0$ ps with two different values of pulse duration ($\tau_{fwhm}$) for (a) $n = 0.05n_c$ and (b) $n = 0.1n_c$. (c), (d) Kinetic energy distribution of electrons at $t = 2.0$ ps for these cases, respectively.}
\label{fig_ex_tau}
\end{figure}

%~~~~~~~~~~~~~~~~~~~~~~~~~~~~~~~~~~~~~~~~~~~~~~~~~~~~~~~~~~~~~~~~~~~
Our study also investigates the impact of laser pulse duration ($\tau_{fwhm}$) on the THz generation process and its efficiency. Figures \ref{fig_thz_tau}(a) and (b) depict the extracted THz field ($E_z^{THz}$) and corresponding Fourier spectra at $x = 700$ $\mu$m i.e., right vacuum of the target, for three different values of $\tau_{fwhm}$ considering photoionizing He target with a fixed electron density plateau $n = 0.05n_c$. Similarly, Fig. \ref{fig_thz_tau}(c) and (d) illustrate the same for $n = 0.1n_c$. For lower density target ($n = 0.05n_c$), THz generation efficiency and, thus, the amplitude of the transmitted THz field increases with a decrease in $\tau_{fwhm}$. This effect is a consequence of the fact that for lower values of $\tau_{fwhm}$, laser ponderomotive force becomes more robust, which amplifies wakefield $E_x$, as shown in Fig. \ref{fig_ex_tau}(a). As a result,  there is an increase in density accumulation at the leading edge of the laser pulse for lower $\tau_{fwhm}$. This effect sharpens the pulse's front edge through local pump depletion, enhancing the transverse drift current. Moreover, the sharper envelope associated with lower $\tau_{fwhm}$ intensifies the asymmetry between the positive and negative half cycles for the photoionization events, consequently leading to a significant increase in the photoionizing current. It is worth noting that when $n = 0.05n_c$ and $\tau_{fwhm} = 100$ fs, the longitudinal length of the laser ($L = c\tau_{fwhm}$) becomes shorter than the plasma wavelength ($\lambda_p$). Consequently, the laser pulse does not undergo self-modulation instability. In this scenario, the dominant processes for THz generation are solely the photoionizing current and the etching of the pulse's leading edge via local pump depletion. Interestingly, for higher density targets ($n = 0.1n_c$), the relationship between the amplitude of the transmitted THz fields and $\tau_{fwhm}$ does not follow a simple monotonic trend, as demonstrated in Figures \ref{fig_thz_tau}(c) and (d). In the case of a higher density target, as the duration of the laser pulse decreases, the pump depletion length ($L_{pd}\propto \tau_{fwhm}/\omega_{pe}$) \cite{lu2007generating} also diminishes. Consequently, after traveling a certain distance within the target, the laser pulse loses most of its energy to the wakefield and accelerated electrons. At a density of $n = 0.1n_c$ and with a pulse duration of $\tau_{fwhm} = 100$ fs, the pump depletion length is approximately $L_{pd} \approx 300$ µm, which is less than the plateau length of the target considered in this study. Thus, for smaller values of $\tau_{fwhm}$, beyond a certain propagation distance ($\sim L_{pd}$), the residual energy of the laser pulse becomes insufficient to effectively drive the wakefield and transverse current, as depicted in Figure \ref{fig_ex_tau}(b). 

%~~~~~~~~~~~~~~~~~~~~~~~~~~~~~~~~~~~~~~~~~~~~~~~~~~~~~~~~~~~~~~~~~~~

%~~~~~~~~~~~~~~~~~~~~~~~~~~~~~~~~~~~~~~~~~~~~~~~~~~~~~~~~~~~~~~~~~~~

Our study further demonstrates that despite using the modest laser power ($\sim 4$ TW), electrons undergo self-injection into the wake structure, leading to significant acceleration to higher energies. The kinetic energy distributions of electrons at $t = 2$ ps with two different values of $\tau_{fwhm}$ have been shown in \ref{fig_ex_tau}(c) and (d) for $n = 0.05n_c$ and $n = 0.1n_c$, respectively. For a low-density target ($n = 0.05n_c$), electrons gained much higher energy for $\tau_{fwhm} = 100$ fs compared to $\tau_{fwhm} = 300$ fs. This outcome results from higher ponderomotive force associated with the sharp laser profile. Conversely, for a higher density target ($n = 0.1n_c$), the maximum energy gained by the electrons is higher for $\tau_{fwhm} = 300$ fs. This effect is attributed to the reduced pump depletion length associated with lower values of $\tau_{fwhm}$ at high-density targets, as discussed earlier. These highly energetic electrons will also emit strong radially polarized THz fields via a coherent transition radiation (CTR) mechanism \cite{PhysRevLett.91.074802, dechard2019thz} while passing through the plasma-vacuum boundary.

%~~~~~~~~~~~~~~~~~~~~~~~~~~~~~~~~~~~~~~~~~~~~~~~~~~~~~~~~~~~~~~~~~~~

%~~~~~~~~~~~~~~~~~~~~~~~~~~~~~~~~~~~~~~~~~~~~~~~~~~~~~~~~~~~~~~~~~~~

\section{Summary and conclusions}
\label{summary}

This study investigates high-field terahertz (THz) generation through the interaction of single-color CO$_2$ laser pulses with helium gas targets. Utilizing multi-dimensional Particle-In-Cell (PIC) simulations, we explore the influence of photoionization on this process by comparing results with the preformed plasma condition. Our findings highlight the significant role of photoionization in the nonlinear laser-matter interaction, particularly in enhancing THz generation efficiency when employing the CO$_2$ laser pulse. We demonstrate a strong dependence of THz radiation generation on target density and laser pulse duration. Through detailed analysis, we disclose the underlying mechanism of THz radiation generation in our study. Specifically, the synergistic coupling of photoionization current, self-modulation instability, and local pump depletion induces a transverse drift current. This current, modulated at the relativistic plasma frequency, emits terahertz (THz) signals polarized parallel to the laser electric field within the frequency range of 3 to 9 THz. Our investigation reveals a favorable parametric regime in which THz fields with peak amplitudes of 150-200 GV/m have been generated and transmitted into the vacuum, which is higher than the previous reports by at least one order of magnitude.

%~~~~~~~~~~~~~~~~~~~~~~~~~~~~~~~~~~~~~~~~~~~~~~~~~~~~~~~~~~~~~~~~~~~

\section*{ACKNOWLEDGMENTS}

This work was supported by the Ministry of Education, Youth and Sports of the Czech Republic through the e-INFRA CZ (ID:90254). The authors also acknowledge the ELI Beamlines HPC facility for computational resources.

%~~~~~~~~~~~~~~~~~~~~~~~~~~~~~~~~~~~~~~~~~~~~~~~~~~~~~~~~~~~~~~~~~~~

\bibliography{ref}

\begin{thebibliography}{83}%
\makeatletter
\providecommand \@ifxundefined [1]{%
 \@ifx{#1\undefined}
}%
\providecommand \@ifnum [1]{%
 \ifnum #1\expandafter \@firstoftwo
 \else \expandafter \@secondoftwo
 \fi
}%
\providecommand \@ifx [1]{%
 \ifx #1\expandafter \@firstoftwo
 \else \expandafter \@secondoftwo
 \fi
}%
\providecommand \natexlab [1]{#1}%
\providecommand \enquote  [1]{``#1''}%
\providecommand \bibnamefont  [1]{#1}%
\providecommand \bibfnamefont [1]{#1}%
\providecommand \citenamefont [1]{#1}%
\providecommand \href@noop [0]{\@secondoftwo}%
\providecommand \href [0]{\begingroup \@sanitize@url \@href}%
\providecommand \@href[1]{\@@startlink{#1}\@@href}%
\providecommand \@@href[1]{\endgroup#1\@@endlink}%
\providecommand \@sanitize@url [0]{\catcode `\\12\catcode `\$12\catcode
  `\&12\catcode `\#12\catcode `\^12\catcode `\_12\catcode `\%12\relax}%
\providecommand \@@startlink[1]{}%
\providecommand \@@endlink[0]{}%
\providecommand \url  [0]{\begingroup\@sanitize@url \@url }%
\providecommand \@url [1]{\endgroup\@href {#1}{\urlprefix }}%
\providecommand \urlprefix  [0]{URL }%
\providecommand \Eprint [0]{\href }%
\providecommand \doibase [0]{http://dx.doi.org/}%
\providecommand \selectlanguage [0]{\@gobble}%
\providecommand \bibinfo  [0]{\@secondoftwo}%
\providecommand \bibfield  [0]{\@secondoftwo}%
\providecommand \translation [1]{[#1]}%
\providecommand \BibitemOpen [0]{}%
\providecommand \bibitemStop [0]{}%
\providecommand \bibitemNoStop [0]{.\EOS\space}%
\providecommand \EOS [0]{\spacefactor3000\relax}%
\providecommand \BibitemShut  [1]{\csname bibitem#1\endcsname}%
\let\auto@bib@innerbib\@empty
%</preamble>
\bibitem [{\citenamefont {Tonouchi}(2007)}]{tonouchi2007cutting}%
  \BibitemOpen
  \bibfield  {author} {\bibinfo {author} {\bibfnamefont {M.}~\bibnamefont
  {Tonouchi}},\ }\href@noop {} {\bibfield  {journal} {\bibinfo  {journal}
  {Nature photonics}\ }\textbf {\bibinfo {volume} {1}},\ \bibinfo {pages} {97}
  (\bibinfo {year} {2007})}\BibitemShut {NoStop}%
\bibitem [{\citenamefont {Kampfrath}\ \emph {et~al.}(2013)\citenamefont
  {Kampfrath}, \citenamefont {Tanaka},\ and\ \citenamefont
  {Nelson}}]{kampfrath2013resonant}%
  \BibitemOpen
  \bibfield  {author} {\bibinfo {author} {\bibfnamefont {T.}~\bibnamefont
  {Kampfrath}}, \bibinfo {author} {\bibfnamefont {K.}~\bibnamefont {Tanaka}}, \
  and\ \bibinfo {author} {\bibfnamefont {K.~A.}\ \bibnamefont {Nelson}},\
  }\href@noop {} {\bibfield  {journal} {\bibinfo  {journal} {Nature Photonics}\
  }\textbf {\bibinfo {volume} {7}},\ \bibinfo {pages} {680} (\bibinfo {year}
  {2013})}\BibitemShut {NoStop}%
\bibitem [{\citenamefont {Chan}\ \emph {et~al.}(2007)\citenamefont {Chan},
  \citenamefont {Deibel},\ and\ \citenamefont {Mittleman}}]{chan2007imaging}%
  \BibitemOpen
  \bibfield  {author} {\bibinfo {author} {\bibfnamefont {W.~L.}\ \bibnamefont
  {Chan}}, \bibinfo {author} {\bibfnamefont {J.}~\bibnamefont {Deibel}}, \ and\
  \bibinfo {author} {\bibfnamefont {D.~M.}\ \bibnamefont {Mittleman}},\
  }\href@noop {} {\bibfield  {journal} {\bibinfo  {journal} {Reports on
  progress in physics}\ }\textbf {\bibinfo {volume} {70}},\ \bibinfo {pages}
  {1325} (\bibinfo {year} {2007})}\BibitemShut {NoStop}%
\bibitem [{\citenamefont {Kemp}\ \emph {et~al.}(2003)\citenamefont {Kemp},
  \citenamefont {Taday}, \citenamefont {Cole}, \citenamefont {Cluff},
  \citenamefont {Fitzgerald},\ and\ \citenamefont {Tribe}}]{kemp2003security}%
  \BibitemOpen
  \bibfield  {author} {\bibinfo {author} {\bibfnamefont {M.~C.}\ \bibnamefont
  {Kemp}}, \bibinfo {author} {\bibfnamefont {P.}~\bibnamefont {Taday}},
  \bibinfo {author} {\bibfnamefont {B.~E.}\ \bibnamefont {Cole}}, \bibinfo
  {author} {\bibfnamefont {J.}~\bibnamefont {Cluff}}, \bibinfo {author}
  {\bibfnamefont {A.~J.}\ \bibnamefont {Fitzgerald}}, \ and\ \bibinfo {author}
  {\bibfnamefont {W.~R.}\ \bibnamefont {Tribe}},\ }in\ \href@noop {} {\emph
  {\bibinfo {booktitle} {Terahertz for military and security applications}}},\
  Vol.\ \bibinfo {volume} {5070}\ (\bibinfo {organization} {SPIE},\ \bibinfo
  {year} {2003})\ pp.\ \bibinfo {pages} {44--52}\BibitemShut {NoStop}%
\bibitem [{\citenamefont {LaRue}\ \emph {et~al.}(2015)\citenamefont {LaRue},
  \citenamefont {Katayama}, \citenamefont {Lindenberg}, \citenamefont {Fisher},
  \citenamefont {{\"O}str{\"o}m}, \citenamefont {Nilsson},\ and\ \citenamefont
  {Ogasawara}}]{larue2015thz}%
  \BibitemOpen
  \bibfield  {author} {\bibinfo {author} {\bibfnamefont {J.~L.}\ \bibnamefont
  {LaRue}}, \bibinfo {author} {\bibfnamefont {T.}~\bibnamefont {Katayama}},
  \bibinfo {author} {\bibfnamefont {A.}~\bibnamefont {Lindenberg}}, \bibinfo
  {author} {\bibfnamefont {A.~S.}\ \bibnamefont {Fisher}}, \bibinfo {author}
  {\bibfnamefont {H.}~\bibnamefont {{\"O}str{\"o}m}}, \bibinfo {author}
  {\bibfnamefont {A.}~\bibnamefont {Nilsson}}, \ and\ \bibinfo {author}
  {\bibfnamefont {H.}~\bibnamefont {Ogasawara}},\ }\href@noop {} {\bibfield
  {journal} {\bibinfo  {journal} {Physical review letters}\ }\textbf {\bibinfo
  {volume} {115}},\ \bibinfo {pages} {036103} (\bibinfo {year}
  {2015})}\BibitemShut {NoStop}%
\bibitem [{\citenamefont {Alexandrov}\ \emph {et~al.}(2013)\citenamefont
  {Alexandrov}, \citenamefont {Phipps}, \citenamefont {Alexandrov},
  \citenamefont {Booshehri}, \citenamefont {Erat}, \citenamefont {Zabolotny},
  \citenamefont {Mielke}, \citenamefont {Chen}, \citenamefont {Rodriguez},
  \citenamefont {Rasmussen} \emph {et~al.}}]{alexandrov2013specificity}%
  \BibitemOpen
  \bibfield  {author} {\bibinfo {author} {\bibfnamefont {B.~S.}\ \bibnamefont
  {Alexandrov}}, \bibinfo {author} {\bibfnamefont {M.~L.}\ \bibnamefont
  {Phipps}}, \bibinfo {author} {\bibfnamefont {L.~B.}\ \bibnamefont
  {Alexandrov}}, \bibinfo {author} {\bibfnamefont {L.~G.}\ \bibnamefont
  {Booshehri}}, \bibinfo {author} {\bibfnamefont {A.}~\bibnamefont {Erat}},
  \bibinfo {author} {\bibfnamefont {J.}~\bibnamefont {Zabolotny}}, \bibinfo
  {author} {\bibfnamefont {C.~H.}\ \bibnamefont {Mielke}}, \bibinfo {author}
  {\bibfnamefont {H.-T.}\ \bibnamefont {Chen}}, \bibinfo {author}
  {\bibfnamefont {G.}~\bibnamefont {Rodriguez}}, \bibinfo {author}
  {\bibfnamefont {K.~{\O}.}\ \bibnamefont {Rasmussen}},  \emph {et~al.},\
  }\href@noop {} {\bibfield  {journal} {\bibinfo  {journal} {Scientific
  reports}\ }\textbf {\bibinfo {volume} {3}},\ \bibinfo {pages} {1184}
  (\bibinfo {year} {2013})}\BibitemShut {NoStop}%
\bibitem [{\citenamefont {Ferguson}\ and\ \citenamefont
  {Zhang}(2002)}]{ferguson2002materials}%
  \BibitemOpen
  \bibfield  {author} {\bibinfo {author} {\bibfnamefont {B.}~\bibnamefont
  {Ferguson}}\ and\ \bibinfo {author} {\bibfnamefont {X.-C.}\ \bibnamefont
  {Zhang}},\ }\href@noop {} {\bibfield  {journal} {\bibinfo  {journal} {Nature
  materials}\ }\textbf {\bibinfo {volume} {1}},\ \bibinfo {pages} {26}
  (\bibinfo {year} {2002})}\BibitemShut {NoStop}%
\bibitem [{\citenamefont {Shalaby}\ and\ \citenamefont
  {Hauri}(2015)}]{shalaby2015demonstration}%
  \BibitemOpen
  \bibfield  {author} {\bibinfo {author} {\bibfnamefont {M.}~\bibnamefont
  {Shalaby}}\ and\ \bibinfo {author} {\bibfnamefont {C.~P.}\ \bibnamefont
  {Hauri}},\ }\href@noop {} {\bibfield  {journal} {\bibinfo  {journal} {Nature
  communications}\ }\textbf {\bibinfo {volume} {6}},\ \bibinfo {pages} {5976}
  (\bibinfo {year} {2015})}\BibitemShut {NoStop}%
\bibitem [{\citenamefont {Vicario}\ \emph {et~al.}(2014)\citenamefont
  {Vicario}, \citenamefont {Ovchinnikov}, \citenamefont {Ashitkov},
  \citenamefont {Agranat}, \citenamefont {Fortov},\ and\ \citenamefont
  {Hauri}}]{vicario2014generation}%
  \BibitemOpen
  \bibfield  {author} {\bibinfo {author} {\bibfnamefont {C.}~\bibnamefont
  {Vicario}}, \bibinfo {author} {\bibfnamefont {A.}~\bibnamefont
  {Ovchinnikov}}, \bibinfo {author} {\bibfnamefont {S.}~\bibnamefont
  {Ashitkov}}, \bibinfo {author} {\bibfnamefont {M.}~\bibnamefont {Agranat}},
  \bibinfo {author} {\bibfnamefont {V.}~\bibnamefont {Fortov}}, \ and\ \bibinfo
  {author} {\bibfnamefont {C.}~\bibnamefont {Hauri}},\ }\href@noop {}
  {\bibfield  {journal} {\bibinfo  {journal} {Optics letters}\ }\textbf
  {\bibinfo {volume} {39}},\ \bibinfo {pages} {6632} (\bibinfo {year}
  {2014})}\BibitemShut {NoStop}%
\bibitem [{\citenamefont {K{\"o}hler}\ \emph {et~al.}(2002)\citenamefont
  {K{\"o}hler}, \citenamefont {Tredicucci}, \citenamefont {Beltram},
  \citenamefont {Beere}, \citenamefont {Linfield}, \citenamefont {Davies},
  \citenamefont {Ritchie}, \citenamefont {Iotti},\ and\ \citenamefont
  {Rossi}}]{kohler2002terahertz}%
  \BibitemOpen
  \bibfield  {author} {\bibinfo {author} {\bibfnamefont {R.}~\bibnamefont
  {K{\"o}hler}}, \bibinfo {author} {\bibfnamefont {A.}~\bibnamefont
  {Tredicucci}}, \bibinfo {author} {\bibfnamefont {F.}~\bibnamefont {Beltram}},
  \bibinfo {author} {\bibfnamefont {H.~E.}\ \bibnamefont {Beere}}, \bibinfo
  {author} {\bibfnamefont {E.~H.}\ \bibnamefont {Linfield}}, \bibinfo {author}
  {\bibfnamefont {A.~G.}\ \bibnamefont {Davies}}, \bibinfo {author}
  {\bibfnamefont {D.~A.}\ \bibnamefont {Ritchie}}, \bibinfo {author}
  {\bibfnamefont {R.~C.}\ \bibnamefont {Iotti}}, \ and\ \bibinfo {author}
  {\bibfnamefont {F.}~\bibnamefont {Rossi}},\ }\href@noop {} {\bibfield
  {journal} {\bibinfo  {journal} {nature}\ }\textbf {\bibinfo {volume} {417}},\
  \bibinfo {pages} {156} (\bibinfo {year} {2002})}\BibitemShut {NoStop}%
\bibitem [{\citenamefont {Chassagneux}\ \emph {et~al.}(2009)\citenamefont
  {Chassagneux}, \citenamefont {Colombelli}, \citenamefont {Maineult},
  \citenamefont {Barbieri}, \citenamefont {Beere}, \citenamefont {Ritchie},
  \citenamefont {Khanna}, \citenamefont {Linfield},\ and\ \citenamefont
  {Davies}}]{chassagneux2009electrically}%
  \BibitemOpen
  \bibfield  {author} {\bibinfo {author} {\bibfnamefont {Y.}~\bibnamefont
  {Chassagneux}}, \bibinfo {author} {\bibfnamefont {R.}~\bibnamefont
  {Colombelli}}, \bibinfo {author} {\bibfnamefont {W.}~\bibnamefont
  {Maineult}}, \bibinfo {author} {\bibfnamefont {S.}~\bibnamefont {Barbieri}},
  \bibinfo {author} {\bibfnamefont {H.}~\bibnamefont {Beere}}, \bibinfo
  {author} {\bibfnamefont {D.}~\bibnamefont {Ritchie}}, \bibinfo {author}
  {\bibfnamefont {S.}~\bibnamefont {Khanna}}, \bibinfo {author} {\bibfnamefont
  {E.}~\bibnamefont {Linfield}}, \ and\ \bibinfo {author} {\bibfnamefont
  {A.~G.}\ \bibnamefont {Davies}},\ }\href@noop {} {\bibfield  {journal}
  {\bibinfo  {journal} {Nature}\ }\textbf {\bibinfo {volume} {457}},\ \bibinfo
  {pages} {174} (\bibinfo {year} {2009})}\BibitemShut {NoStop}%
\bibitem [{\citenamefont {Wu}\ \emph {et~al.}(2013)\citenamefont {Wu},
  \citenamefont {Fisher}, \citenamefont {Goodfellow}, \citenamefont {Fuchs},
  \citenamefont {Daranciang}, \citenamefont {Hogan}, \citenamefont {Loos},\
  and\ \citenamefont {Lindenberg}}]{wu2013intense}%
  \BibitemOpen
  \bibfield  {author} {\bibinfo {author} {\bibfnamefont {Z.}~\bibnamefont
  {Wu}}, \bibinfo {author} {\bibfnamefont {A.~S.}\ \bibnamefont {Fisher}},
  \bibinfo {author} {\bibfnamefont {J.}~\bibnamefont {Goodfellow}}, \bibinfo
  {author} {\bibfnamefont {M.}~\bibnamefont {Fuchs}}, \bibinfo {author}
  {\bibfnamefont {D.}~\bibnamefont {Daranciang}}, \bibinfo {author}
  {\bibfnamefont {M.}~\bibnamefont {Hogan}}, \bibinfo {author} {\bibfnamefont
  {H.}~\bibnamefont {Loos}}, \ and\ \bibinfo {author} {\bibfnamefont
  {A.}~\bibnamefont {Lindenberg}},\ }\href@noop {} {\bibfield  {journal}
  {\bibinfo  {journal} {Review of Scientific Instruments}\ }\textbf {\bibinfo
  {volume} {84}} (\bibinfo {year} {2013})}\BibitemShut {NoStop}%
\bibitem [{\citenamefont {Ramian}(1992)}]{ramian1992new}%
  \BibitemOpen
  \bibfield  {author} {\bibinfo {author} {\bibfnamefont {G.}~\bibnamefont
  {Ramian}},\ }\href@noop {} {\bibfield  {journal} {\bibinfo  {journal}
  {Nuclear Instruments and Methods in Physics Research Section A: Accelerators,
  Spectrometers, Detectors and Associated Equipment}\ }\textbf {\bibinfo
  {volume} {318}},\ \bibinfo {pages} {225} (\bibinfo {year}
  {1992})}\BibitemShut {NoStop}%
\bibitem [{\citenamefont {Cook}\ and\ \citenamefont
  {Hochstrasser}(2000)}]{cook2000intense}%
  \BibitemOpen
  \bibfield  {author} {\bibinfo {author} {\bibfnamefont {D.}~\bibnamefont
  {Cook}}\ and\ \bibinfo {author} {\bibfnamefont {R.}~\bibnamefont
  {Hochstrasser}},\ }\href@noop {} {\bibfield  {journal} {\bibinfo  {journal}
  {Optics letters}\ }\textbf {\bibinfo {volume} {25}},\ \bibinfo {pages} {1210}
  (\bibinfo {year} {2000})}\BibitemShut {NoStop}%
\bibitem [{\citenamefont {Clerici}\ \emph {et~al.}(2013)\citenamefont
  {Clerici}, \citenamefont {Peccianti}, \citenamefont {Schmidt}, \citenamefont
  {Caspani}, \citenamefont {Shalaby}, \citenamefont {Giguere}, \citenamefont
  {Lotti}, \citenamefont {Couairon}, \citenamefont {L{\'e}gar{\'e}},
  \citenamefont {Ozaki} \emph {et~al.}}]{clerici2013wavelength}%
  \BibitemOpen
  \bibfield  {author} {\bibinfo {author} {\bibfnamefont {M.}~\bibnamefont
  {Clerici}}, \bibinfo {author} {\bibfnamefont {M.}~\bibnamefont {Peccianti}},
  \bibinfo {author} {\bibfnamefont {B.~E.}\ \bibnamefont {Schmidt}}, \bibinfo
  {author} {\bibfnamefont {L.}~\bibnamefont {Caspani}}, \bibinfo {author}
  {\bibfnamefont {M.}~\bibnamefont {Shalaby}}, \bibinfo {author} {\bibfnamefont
  {M.}~\bibnamefont {Giguere}}, \bibinfo {author} {\bibfnamefont
  {A.}~\bibnamefont {Lotti}}, \bibinfo {author} {\bibfnamefont
  {A.}~\bibnamefont {Couairon}}, \bibinfo {author} {\bibfnamefont
  {F.}~\bibnamefont {L{\'e}gar{\'e}}}, \bibinfo {author} {\bibfnamefont
  {T.}~\bibnamefont {Ozaki}},  \emph {et~al.},\ }\href@noop {} {\bibfield
  {journal} {\bibinfo  {journal} {Physical Review Letters}\ }\textbf {\bibinfo
  {volume} {110}},\ \bibinfo {pages} {253901} (\bibinfo {year}
  {2013})}\BibitemShut {NoStop}%
\bibitem [{\citenamefont {Andreeva}\ \emph {et~al.}(2016)\citenamefont
  {Andreeva}, \citenamefont {Kosareva}, \citenamefont {Panov}, \citenamefont
  {Shipilo}, \citenamefont {Solyankin}, \citenamefont {Esaulkov}, \citenamefont
  {de~Alaiza~Mart{\'\i}nez}, \citenamefont {Shkurinov}, \citenamefont
  {Makarov}, \citenamefont {Berg{\'e}} \emph
  {et~al.}}]{andreeva2016ultrabroad}%
  \BibitemOpen
  \bibfield  {author} {\bibinfo {author} {\bibfnamefont {V.}~\bibnamefont
  {Andreeva}}, \bibinfo {author} {\bibfnamefont {O.}~\bibnamefont {Kosareva}},
  \bibinfo {author} {\bibfnamefont {N.}~\bibnamefont {Panov}}, \bibinfo
  {author} {\bibfnamefont {D.}~\bibnamefont {Shipilo}}, \bibinfo {author}
  {\bibfnamefont {P.}~\bibnamefont {Solyankin}}, \bibinfo {author}
  {\bibfnamefont {M.}~\bibnamefont {Esaulkov}}, \bibinfo {author}
  {\bibfnamefont {P.~G.}\ \bibnamefont {de~Alaiza~Mart{\'\i}nez}}, \bibinfo
  {author} {\bibfnamefont {A.}~\bibnamefont {Shkurinov}}, \bibinfo {author}
  {\bibfnamefont {V.}~\bibnamefont {Makarov}}, \bibinfo {author} {\bibfnamefont
  {L.}~\bibnamefont {Berg{\'e}}},  \emph {et~al.},\ }\href@noop {} {\bibfield
  {journal} {\bibinfo  {journal} {Physical review letters}\ }\textbf {\bibinfo
  {volume} {116}},\ \bibinfo {pages} {063902} (\bibinfo {year}
  {2016})}\BibitemShut {NoStop}%
\bibitem [{\citenamefont {Kim}\ \emph {et~al.}(2008)\citenamefont {Kim},
  \citenamefont {Taylor}, \citenamefont {Glownia},\ and\ \citenamefont
  {Rodriguez}}]{kim2008coherent}%
  \BibitemOpen
  \bibfield  {author} {\bibinfo {author} {\bibfnamefont {K.-Y.}\ \bibnamefont
  {Kim}}, \bibinfo {author} {\bibfnamefont {A.}~\bibnamefont {Taylor}},
  \bibinfo {author} {\bibfnamefont {J.}~\bibnamefont {Glownia}}, \ and\
  \bibinfo {author} {\bibfnamefont {G.}~\bibnamefont {Rodriguez}},\ }\href@noop
  {} {\bibfield  {journal} {\bibinfo  {journal} {Nature photonics}\ }\textbf
  {\bibinfo {volume} {2}},\ \bibinfo {pages} {605} (\bibinfo {year}
  {2008})}\BibitemShut {NoStop}%
\bibitem [{\citenamefont {Tajima}\ and\ \citenamefont
  {Dawson}(1979)}]{tajima1979laser}%
  \BibitemOpen
  \bibfield  {author} {\bibinfo {author} {\bibfnamefont {T.}~\bibnamefont
  {Tajima}}\ and\ \bibinfo {author} {\bibfnamefont {J.~M.}\ \bibnamefont
  {Dawson}},\ }\href@noop {} {\bibfield  {journal} {\bibinfo  {journal}
  {Physical review letters}\ }\textbf {\bibinfo {volume} {43}},\ \bibinfo
  {pages} {267} (\bibinfo {year} {1979})}\BibitemShut {NoStop}%
\bibitem [{\citenamefont {Joshi}(2007)}]{joshi2007development}%
  \BibitemOpen
  \bibfield  {author} {\bibinfo {author} {\bibfnamefont {C.}~\bibnamefont
  {Joshi}},\ }\href@noop {} {\bibfield  {journal} {\bibinfo  {journal} {Physics
  of plasmas}\ }\textbf {\bibinfo {volume} {14}} (\bibinfo {year}
  {2007})}\BibitemShut {NoStop}%
\bibitem [{\citenamefont {Esarey}\ \emph {et~al.}(2009)\citenamefont {Esarey},
  \citenamefont {Schroeder},\ and\ \citenamefont
  {Leemans}}]{RevModPhys.81.1229}%
  \BibitemOpen
  \bibfield  {author} {\bibinfo {author} {\bibfnamefont {E.}~\bibnamefont
  {Esarey}}, \bibinfo {author} {\bibfnamefont {C.~B.}\ \bibnamefont
  {Schroeder}}, \ and\ \bibinfo {author} {\bibfnamefont {W.~P.}\ \bibnamefont
  {Leemans}},\ }\href {\doibase 10.1103/RevModPhys.81.1229} {\bibfield
  {journal} {\bibinfo  {journal} {Rev. Mod. Phys.}\ }\textbf {\bibinfo {volume}
  {81}},\ \bibinfo {pages} {1229} (\bibinfo {year} {2009})}\BibitemShut
  {NoStop}%
\bibitem [{\citenamefont {Wen}\ \emph {et~al.}(2020)\citenamefont {Wen},
  \citenamefont {Salamin},\ and\ \citenamefont
  {Keitel}}]{PhysRevApplied.13.034001}%
  \BibitemOpen
  \bibfield  {author} {\bibinfo {author} {\bibfnamefont {M.}~\bibnamefont
  {Wen}}, \bibinfo {author} {\bibfnamefont {Y.~I.}\ \bibnamefont {Salamin}}, \
  and\ \bibinfo {author} {\bibfnamefont {C.~H.}\ \bibnamefont {Keitel}},\
  }\href {\doibase 10.1103/PhysRevApplied.13.034001} {\bibfield  {journal}
  {\bibinfo  {journal} {Phys. Rev. Appl.}\ }\textbf {\bibinfo {volume} {13}},\
  \bibinfo {pages} {034001} (\bibinfo {year} {2020})}\BibitemShut {NoStop}%
\bibitem [{\citenamefont {Shen}\ \emph {et~al.}(2022)\citenamefont {Shen},
  \citenamefont {Pukhov}, \citenamefont {Rosmej},\ and\ \citenamefont
  {Andreev}}]{PhysRevApplied.18.064091}%
  \BibitemOpen
  \bibfield  {author} {\bibinfo {author} {\bibfnamefont {X.~F.}\ \bibnamefont
  {Shen}}, \bibinfo {author} {\bibfnamefont {A.}~\bibnamefont {Pukhov}},
  \bibinfo {author} {\bibfnamefont {O.~N.}\ \bibnamefont {Rosmej}}, \ and\
  \bibinfo {author} {\bibfnamefont {N.~E.}\ \bibnamefont {Andreev}},\ }\href
  {\doibase 10.1103/PhysRevApplied.18.064091} {\bibfield  {journal} {\bibinfo
  {journal} {Phys. Rev. Appl.}\ }\textbf {\bibinfo {volume} {18}},\ \bibinfo
  {pages} {064091} (\bibinfo {year} {2022})}\BibitemShut {NoStop}%
\bibitem [{\citenamefont {Zhu}\ \emph {et~al.}(2021)\citenamefont {Zhu},
  \citenamefont {Liu}, \citenamefont {Chen}, \citenamefont {Weng},
  \citenamefont {He}, \citenamefont {Assmann}, \citenamefont {Sheng},\ and\
  \citenamefont {Zhang}}]{PhysRevApplied.15.044039}%
  \BibitemOpen
  \bibfield  {author} {\bibinfo {author} {\bibfnamefont {X.-L.}\ \bibnamefont
  {Zhu}}, \bibinfo {author} {\bibfnamefont {W.-Y.}\ \bibnamefont {Liu}},
  \bibinfo {author} {\bibfnamefont {M.}~\bibnamefont {Chen}}, \bibinfo {author}
  {\bibfnamefont {S.-M.}\ \bibnamefont {Weng}}, \bibinfo {author}
  {\bibfnamefont {F.}~\bibnamefont {He}}, \bibinfo {author} {\bibfnamefont
  {R.}~\bibnamefont {Assmann}}, \bibinfo {author} {\bibfnamefont {Z.-M.}\
  \bibnamefont {Sheng}}, \ and\ \bibinfo {author} {\bibfnamefont
  {J.}~\bibnamefont {Zhang}},\ }\href {\doibase
  10.1103/PhysRevApplied.15.044039} {\bibfield  {journal} {\bibinfo  {journal}
  {Phys. Rev. Appl.}\ }\textbf {\bibinfo {volume} {15}},\ \bibinfo {pages}
  {044039} (\bibinfo {year} {2021})}\BibitemShut {NoStop}%
\bibitem [{\citenamefont {Maity}\ \emph {et~al.}(2024)\citenamefont {Maity},
  \citenamefont {Mondal}, \citenamefont {Vishnyakov},\ and\ \citenamefont
  {Molodozhentsev}}]{maity2024parametric}%
  \BibitemOpen
  \bibfield  {author} {\bibinfo {author} {\bibfnamefont {S.}~\bibnamefont
  {Maity}}, \bibinfo {author} {\bibfnamefont {A.}~\bibnamefont {Mondal}},
  \bibinfo {author} {\bibfnamefont {E.}~\bibnamefont {Vishnyakov}}, \ and\
  \bibinfo {author} {\bibfnamefont {A.}~\bibnamefont {Molodozhentsev}},\
  }\href@noop {} {\bibfield  {journal} {\bibinfo  {journal} {Plasma Physics and
  Controlled Fusion}\ }\textbf {\bibinfo {volume} {66}},\ \bibinfo {pages}
  {035012} (\bibinfo {year} {2024})}\BibitemShut {NoStop}%
\bibitem [{\citenamefont {Macchi}\ \emph {et~al.}(2013)\citenamefont {Macchi},
  \citenamefont {Borghesi},\ and\ \citenamefont {Passoni}}]{macchi2013ion}%
  \BibitemOpen
  \bibfield  {author} {\bibinfo {author} {\bibfnamefont {A.}~\bibnamefont
  {Macchi}}, \bibinfo {author} {\bibfnamefont {M.}~\bibnamefont {Borghesi}}, \
  and\ \bibinfo {author} {\bibfnamefont {M.}~\bibnamefont {Passoni}},\
  }\href@noop {} {\bibfield  {journal} {\bibinfo  {journal} {Reviews of Modern
  Physics}\ }\textbf {\bibinfo {volume} {85}},\ \bibinfo {pages} {751}
  (\bibinfo {year} {2013})}\BibitemShut {NoStop}%
\bibitem [{\citenamefont {Daido}\ \emph {et~al.}(2012)\citenamefont {Daido},
  \citenamefont {Nishiuchi},\ and\ \citenamefont
  {Pirozhkov}}]{daido2012review}%
  \BibitemOpen
  \bibfield  {author} {\bibinfo {author} {\bibfnamefont {H.}~\bibnamefont
  {Daido}}, \bibinfo {author} {\bibfnamefont {M.}~\bibnamefont {Nishiuchi}}, \
  and\ \bibinfo {author} {\bibfnamefont {A.~S.}\ \bibnamefont {Pirozhkov}},\
  }\href@noop {} {\bibfield  {journal} {\bibinfo  {journal} {Reports on
  progress in physics}\ }\textbf {\bibinfo {volume} {75}},\ \bibinfo {pages}
  {056401} (\bibinfo {year} {2012})}\BibitemShut {NoStop}%
\bibitem [{\citenamefont {Zhang}\ \emph {et~al.}(2021)\citenamefont {Zhang},
  \citenamefont {Zhong}, \citenamefont {Zhu}, \citenamefont {He}, \citenamefont
  {Zepf},\ and\ \citenamefont {Qiao}}]{PhysRevApplied.16.024042}%
  \BibitemOpen
  \bibfield  {author} {\bibinfo {author} {\bibfnamefont {Y.}~\bibnamefont
  {Zhang}}, \bibinfo {author} {\bibfnamefont {C.~L.}\ \bibnamefont {Zhong}},
  \bibinfo {author} {\bibfnamefont {S.~P.}\ \bibnamefont {Zhu}}, \bibinfo
  {author} {\bibfnamefont {X.~T.}\ \bibnamefont {He}}, \bibinfo {author}
  {\bibfnamefont {M.}~\bibnamefont {Zepf}}, \ and\ \bibinfo {author}
  {\bibfnamefont {B.}~\bibnamefont {Qiao}},\ }\href {\doibase
  10.1103/PhysRevApplied.16.024042} {\bibfield  {journal} {\bibinfo  {journal}
  {Phys. Rev. Appl.}\ }\textbf {\bibinfo {volume} {16}},\ \bibinfo {pages}
  {024042} (\bibinfo {year} {2021})}\BibitemShut {NoStop}%
\bibitem [{\citenamefont {Ganeev}(2007)}]{ganeev2007high}%
  \BibitemOpen
  \bibfield  {author} {\bibinfo {author} {\bibfnamefont {R.}~\bibnamefont
  {Ganeev}},\ }\href@noop {} {\bibfield  {journal} {\bibinfo  {journal}
  {Journal of Physics B: Atomic, Molecular and Optical Physics}\ }\textbf
  {\bibinfo {volume} {40}},\ \bibinfo {pages} {R213} (\bibinfo {year}
  {2007})}\BibitemShut {NoStop}%
\bibitem [{\citenamefont {Ganeev}\ \emph {et~al.}(2009)\citenamefont {Ganeev},
  \citenamefont {Bom}, \citenamefont {Abdul-Hadi}, \citenamefont {Wong},
  \citenamefont {Brichta}, \citenamefont {Bhardwaj},\ and\ \citenamefont
  {Ozaki}}]{ganeev2009higher}%
  \BibitemOpen
  \bibfield  {author} {\bibinfo {author} {\bibfnamefont {R.}~\bibnamefont
  {Ganeev}}, \bibinfo {author} {\bibfnamefont {L.~E.}\ \bibnamefont {Bom}},
  \bibinfo {author} {\bibfnamefont {J.}~\bibnamefont {Abdul-Hadi}}, \bibinfo
  {author} {\bibfnamefont {M.}~\bibnamefont {Wong}}, \bibinfo {author}
  {\bibfnamefont {J.}~\bibnamefont {Brichta}}, \bibinfo {author} {\bibfnamefont
  {V.}~\bibnamefont {Bhardwaj}}, \ and\ \bibinfo {author} {\bibfnamefont
  {T.}~\bibnamefont {Ozaki}},\ }\href@noop {} {\bibfield  {journal} {\bibinfo
  {journal} {Physical Review Letters}\ }\textbf {\bibinfo {volume} {102}},\
  \bibinfo {pages} {013903} (\bibinfo {year} {2009})}\BibitemShut {NoStop}%
\bibitem [{\citenamefont {Hu}\ \emph {et~al.}(2010)\citenamefont {Hu},
  \citenamefont {Militzer}, \citenamefont {Goncharov},\ and\ \citenamefont
  {Skupsky}}]{hu2010strong}%
  \BibitemOpen
  \bibfield  {author} {\bibinfo {author} {\bibfnamefont {S.}~\bibnamefont
  {Hu}}, \bibinfo {author} {\bibfnamefont {B.}~\bibnamefont {Militzer}},
  \bibinfo {author} {\bibfnamefont {V.}~\bibnamefont {Goncharov}}, \ and\
  \bibinfo {author} {\bibfnamefont {S.}~\bibnamefont {Skupsky}},\ }\href@noop
  {} {\bibfield  {journal} {\bibinfo  {journal} {Physical review letters}\
  }\textbf {\bibinfo {volume} {104}},\ \bibinfo {pages} {235003} (\bibinfo
  {year} {2010})}\BibitemShut {NoStop}%
\bibitem [{\citenamefont {Zylstra}\ \emph {et~al.}(2021)\citenamefont
  {Zylstra}, \citenamefont {Kritcher}, \citenamefont {Hurricane}, \citenamefont
  {Callahan}, \citenamefont {Baker}, \citenamefont {Braun}, \citenamefont
  {Casey}, \citenamefont {Clark}, \citenamefont {Clark}, \citenamefont
  {D{\"o}ppner} \emph {et~al.}}]{zylstra2021record}%
  \BibitemOpen
  \bibfield  {author} {\bibinfo {author} {\bibfnamefont {A.}~\bibnamefont
  {Zylstra}}, \bibinfo {author} {\bibfnamefont {A.}~\bibnamefont {Kritcher}},
  \bibinfo {author} {\bibfnamefont {O.}~\bibnamefont {Hurricane}}, \bibinfo
  {author} {\bibfnamefont {D.}~\bibnamefont {Callahan}}, \bibinfo {author}
  {\bibfnamefont {K.}~\bibnamefont {Baker}}, \bibinfo {author} {\bibfnamefont
  {T.}~\bibnamefont {Braun}}, \bibinfo {author} {\bibfnamefont
  {D.}~\bibnamefont {Casey}}, \bibinfo {author} {\bibfnamefont
  {D.}~\bibnamefont {Clark}}, \bibinfo {author} {\bibfnamefont
  {K.}~\bibnamefont {Clark}}, \bibinfo {author} {\bibfnamefont
  {T.}~\bibnamefont {D{\"o}ppner}},  \emph {et~al.},\ }\href@noop {} {\bibfield
   {journal} {\bibinfo  {journal} {Physical review letters}\ }\textbf {\bibinfo
  {volume} {126}},\ \bibinfo {pages} {025001} (\bibinfo {year}
  {2021})}\BibitemShut {NoStop}%
\bibitem [{\citenamefont {Igumenshchev}\ \emph {et~al.}(2023)\citenamefont
  {Igumenshchev}, \citenamefont {Theobald}, \citenamefont {Stoeckl},
  \citenamefont {Shah}, \citenamefont {Bishel}, \citenamefont {Goncharov},
  \citenamefont {Bonino}, \citenamefont {Campbell}, \citenamefont {Ceurvorst},
  \citenamefont {Chin} \emph {et~al.}}]{igumenshchev2023proof}%
  \BibitemOpen
  \bibfield  {author} {\bibinfo {author} {\bibfnamefont {I.}~\bibnamefont
  {Igumenshchev}}, \bibinfo {author} {\bibfnamefont {W.}~\bibnamefont
  {Theobald}}, \bibinfo {author} {\bibfnamefont {C.}~\bibnamefont {Stoeckl}},
  \bibinfo {author} {\bibfnamefont {R.}~\bibnamefont {Shah}}, \bibinfo {author}
  {\bibfnamefont {D.}~\bibnamefont {Bishel}}, \bibinfo {author} {\bibfnamefont
  {V.}~\bibnamefont {Goncharov}}, \bibinfo {author} {\bibfnamefont
  {M.}~\bibnamefont {Bonino}}, \bibinfo {author} {\bibfnamefont
  {E.}~\bibnamefont {Campbell}}, \bibinfo {author} {\bibfnamefont
  {L.}~\bibnamefont {Ceurvorst}}, \bibinfo {author} {\bibfnamefont
  {D.}~\bibnamefont {Chin}},  \emph {et~al.},\ }\href@noop {} {\bibfield
  {journal} {\bibinfo  {journal} {Physical Review Letters}\ }\textbf {\bibinfo
  {volume} {131}},\ \bibinfo {pages} {015102} (\bibinfo {year}
  {2023})}\BibitemShut {NoStop}%
\bibitem [{\citenamefont {Hamster}\ \emph {et~al.}(1993)\citenamefont
  {Hamster}, \citenamefont {Sullivan}, \citenamefont {Gordon}, \citenamefont
  {White},\ and\ \citenamefont {Falcone}}]{hamster1993subpicosecond}%
  \BibitemOpen
  \bibfield  {author} {\bibinfo {author} {\bibfnamefont {H.}~\bibnamefont
  {Hamster}}, \bibinfo {author} {\bibfnamefont {A.}~\bibnamefont {Sullivan}},
  \bibinfo {author} {\bibfnamefont {S.}~\bibnamefont {Gordon}}, \bibinfo
  {author} {\bibfnamefont {W.}~\bibnamefont {White}}, \ and\ \bibinfo {author}
  {\bibfnamefont {R.}~\bibnamefont {Falcone}},\ }\href@noop {} {\bibfield
  {journal} {\bibinfo  {journal} {Physical review letters}\ }\textbf {\bibinfo
  {volume} {71}},\ \bibinfo {pages} {2725} (\bibinfo {year}
  {1993})}\BibitemShut {NoStop}%
\bibitem [{\citenamefont {Kim}\ \emph {et~al.}(2007)\citenamefont {Kim},
  \citenamefont {Glownia}, \citenamefont {Taylor},\ and\ \citenamefont
  {Rodriguez}}]{kim2007terahertz}%
  \BibitemOpen
  \bibfield  {author} {\bibinfo {author} {\bibfnamefont {K.-Y.}\ \bibnamefont
  {Kim}}, \bibinfo {author} {\bibfnamefont {J.~H.}\ \bibnamefont {Glownia}},
  \bibinfo {author} {\bibfnamefont {A.~J.}\ \bibnamefont {Taylor}}, \ and\
  \bibinfo {author} {\bibfnamefont {G.}~\bibnamefont {Rodriguez}},\ }\href@noop
  {} {\bibfield  {journal} {\bibinfo  {journal} {Optics express}\ }\textbf
  {\bibinfo {volume} {15}},\ \bibinfo {pages} {4577} (\bibinfo {year}
  {2007})}\BibitemShut {NoStop}%
\bibitem [{\citenamefont {Sheng}\ \emph {et~al.}(2005)\citenamefont {Sheng},
  \citenamefont {Mima}, \citenamefont {Zhang},\ and\ \citenamefont
  {Sanuki}}]{sheng2005emission}%
  \BibitemOpen
  \bibfield  {author} {\bibinfo {author} {\bibfnamefont {Z.-M.}\ \bibnamefont
  {Sheng}}, \bibinfo {author} {\bibfnamefont {K.}~\bibnamefont {Mima}},
  \bibinfo {author} {\bibfnamefont {J.}~\bibnamefont {Zhang}}, \ and\ \bibinfo
  {author} {\bibfnamefont {H.}~\bibnamefont {Sanuki}},\ }\href@noop {}
  {\bibfield  {journal} {\bibinfo  {journal} {Physical review letters}\
  }\textbf {\bibinfo {volume} {94}},\ \bibinfo {pages} {095003} (\bibinfo
  {year} {2005})}\BibitemShut {NoStop}%
\bibitem [{\citenamefont {Wang}\ \emph {et~al.}(2024)\citenamefont {Wang},
  \citenamefont {Zhang}, \citenamefont {Chen}, \citenamefont {Chen},
  \citenamefont {Hu}, \citenamefont {Zhu}, \citenamefont {Yan}, \citenamefont
  {Xu}, \citenamefont {Sun}, \citenamefont {Chen}, \citenamefont {Liu},
  \citenamefont {Chen}, \citenamefont {Zhang},\ and\ \citenamefont
  {Sheng}}]{PhysRevLett.132.165002}%
  \BibitemOpen
  \bibfield  {author} {\bibinfo {author} {\bibfnamefont {L.}~\bibnamefont
  {Wang}}, \bibinfo {author} {\bibfnamefont {Z.}~\bibnamefont {Zhang}},
  \bibinfo {author} {\bibfnamefont {S.}~\bibnamefont {Chen}}, \bibinfo {author}
  {\bibfnamefont {Y.}~\bibnamefont {Chen}}, \bibinfo {author} {\bibfnamefont
  {X.}~\bibnamefont {Hu}}, \bibinfo {author} {\bibfnamefont {M.}~\bibnamefont
  {Zhu}}, \bibinfo {author} {\bibfnamefont {W.}~\bibnamefont {Yan}}, \bibinfo
  {author} {\bibfnamefont {H.}~\bibnamefont {Xu}}, \bibinfo {author}
  {\bibfnamefont {L.}~\bibnamefont {Sun}}, \bibinfo {author} {\bibfnamefont
  {M.}~\bibnamefont {Chen}}, \bibinfo {author} {\bibfnamefont {F.}~\bibnamefont
  {Liu}}, \bibinfo {author} {\bibfnamefont {L.}~\bibnamefont {Chen}}, \bibinfo
  {author} {\bibfnamefont {J.}~\bibnamefont {Zhang}}, \ and\ \bibinfo {author}
  {\bibfnamefont {Z.}~\bibnamefont {Sheng}},\ }\href {\doibase
  10.1103/PhysRevLett.132.165002} {\bibfield  {journal} {\bibinfo  {journal}
  {Phys. Rev. Lett.}\ }\textbf {\bibinfo {volume} {132}},\ \bibinfo {pages}
  {165002} (\bibinfo {year} {2024})}\BibitemShut {NoStop}%
\bibitem [{\citenamefont {Chen}\ and\ \citenamefont
  {Pukhov}(2015)}]{chen2015high}%
  \BibitemOpen
  \bibfield  {author} {\bibinfo {author} {\bibfnamefont {Z.-Y.}\ \bibnamefont
  {Chen}}\ and\ \bibinfo {author} {\bibfnamefont {A.}~\bibnamefont {Pukhov}},\
  }\href {\doibase 10.1063/1.4933130} {\bibfield  {journal} {\bibinfo
  {journal} {Physics of Plasmas}\ }\textbf {\bibinfo {volume} {22}} (\bibinfo
  {year} {2015}),\ 10.1063/1.4933130}\BibitemShut {NoStop}%
\bibitem [{\citenamefont {Decker}\ \emph {et~al.}(1996)\citenamefont {Decker},
  \citenamefont {Mori}, \citenamefont {Tzeng},\ and\ \citenamefont
  {Katsouleas}}]{decker1996evolution}%
  \BibitemOpen
  \bibfield  {author} {\bibinfo {author} {\bibfnamefont {C.}~\bibnamefont
  {Decker}}, \bibinfo {author} {\bibfnamefont {W.}~\bibnamefont {Mori}},
  \bibinfo {author} {\bibfnamefont {K.-C.}\ \bibnamefont {Tzeng}}, \ and\
  \bibinfo {author} {\bibfnamefont {T.}~\bibnamefont {Katsouleas}},\ }\href
  {\doibase 10.1063/1.872001} {\bibfield  {journal} {\bibinfo  {journal}
  {Physics of Plasmas}\ }\textbf {\bibinfo {volume} {3}},\ \bibinfo {pages}
  {2047} (\bibinfo {year} {1996})}\BibitemShut {NoStop}%
\bibitem [{\citenamefont {Wang}\ \emph {et~al.}(2015)\citenamefont {Wang},
  \citenamefont {Gibbon}, \citenamefont {Sheng},\ and\ \citenamefont
  {Li}}]{wang2015tunable}%
  \BibitemOpen
  \bibfield  {author} {\bibinfo {author} {\bibfnamefont {W.-M.}\ \bibnamefont
  {Wang}}, \bibinfo {author} {\bibfnamefont {P.}~\bibnamefont {Gibbon}},
  \bibinfo {author} {\bibfnamefont {Z.-M.}\ \bibnamefont {Sheng}}, \ and\
  \bibinfo {author} {\bibfnamefont {Y.-T.}\ \bibnamefont {Li}},\ }\href@noop {}
  {\bibfield  {journal} {\bibinfo  {journal} {Physical review letters}\
  }\textbf {\bibinfo {volume} {114}},\ \bibinfo {pages} {253901} (\bibinfo
  {year} {2015})}\BibitemShut {NoStop}%
\bibitem [{\citenamefont {Tailliez}\ \emph {et~al.}(2022)\citenamefont
  {Tailliez}, \citenamefont {Davoine}, \citenamefont {Debayle}, \citenamefont
  {Gremillet},\ and\ \citenamefont {Berg{\'e}}}]{tailliez2022terahertz}%
  \BibitemOpen
  \bibfield  {author} {\bibinfo {author} {\bibfnamefont {C.}~\bibnamefont
  {Tailliez}}, \bibinfo {author} {\bibfnamefont {X.}~\bibnamefont {Davoine}},
  \bibinfo {author} {\bibfnamefont {A.}~\bibnamefont {Debayle}}, \bibinfo
  {author} {\bibfnamefont {L.}~\bibnamefont {Gremillet}}, \ and\ \bibinfo
  {author} {\bibfnamefont {L.}~\bibnamefont {Berg{\'e}}},\ }\href@noop {}
  {\bibfield  {journal} {\bibinfo  {journal} {Physical Review Letters}\
  }\textbf {\bibinfo {volume} {128}},\ \bibinfo {pages} {174802} (\bibinfo
  {year} {2022})}\BibitemShut {NoStop}%
\bibitem [{\citenamefont {Koulouklidis}\ \emph {et~al.}(2020)\citenamefont
  {Koulouklidis}, \citenamefont {Gollner}, \citenamefont {Shumakova},
  \citenamefont {Fedorov}, \citenamefont {Pug{\v{z}}lys}, \citenamefont
  {Baltu{\v{s}}ka},\ and\ \citenamefont
  {Tzortzakis}}]{koulouklidis2020observation}%
  \BibitemOpen
  \bibfield  {author} {\bibinfo {author} {\bibfnamefont {A.~D.}\ \bibnamefont
  {Koulouklidis}}, \bibinfo {author} {\bibfnamefont {C.}~\bibnamefont
  {Gollner}}, \bibinfo {author} {\bibfnamefont {V.}~\bibnamefont {Shumakova}},
  \bibinfo {author} {\bibfnamefont {V.~Y.}\ \bibnamefont {Fedorov}}, \bibinfo
  {author} {\bibfnamefont {A.}~\bibnamefont {Pug{\v{z}}lys}}, \bibinfo {author}
  {\bibfnamefont {A.}~\bibnamefont {Baltu{\v{s}}ka}}, \ and\ \bibinfo {author}
  {\bibfnamefont {S.}~\bibnamefont {Tzortzakis}},\ }\href@noop {} {\bibfield
  {journal} {\bibinfo  {journal} {Nature communications}\ }\textbf {\bibinfo
  {volume} {11}},\ \bibinfo {pages} {292} (\bibinfo {year} {2020})}\BibitemShut
  {NoStop}%
\bibitem [{\citenamefont {Nguyen}\ \emph {et~al.}(2018)\citenamefont {Nguyen},
  \citenamefont {de~Alaiza~Mart{\'\i}nez}, \citenamefont {Thiele},
  \citenamefont {Skupin},\ and\ \citenamefont
  {Berg{\'e}}}]{nguyen2018broadband}%
  \BibitemOpen
  \bibfield  {author} {\bibinfo {author} {\bibfnamefont {A.}~\bibnamefont
  {Nguyen}}, \bibinfo {author} {\bibfnamefont {P.~G.}\ \bibnamefont
  {de~Alaiza~Mart{\'\i}nez}}, \bibinfo {author} {\bibfnamefont
  {I.}~\bibnamefont {Thiele}}, \bibinfo {author} {\bibfnamefont
  {S.}~\bibnamefont {Skupin}}, \ and\ \bibinfo {author} {\bibfnamefont
  {L.}~\bibnamefont {Berg{\'e}}},\ }\href@noop {} {\bibfield  {journal}
  {\bibinfo  {journal} {Physical Review A}\ }\textbf {\bibinfo {volume} {97}},\
  \bibinfo {pages} {063839} (\bibinfo {year} {2018})}\BibitemShut {NoStop}%
\bibitem [{\citenamefont {Wang}\ \emph {et~al.}(2011)\citenamefont {Wang},
  \citenamefont {Kawata}, \citenamefont {Sheng}, \citenamefont {Li},
  \citenamefont {Chen}, \citenamefont {Qian},\ and\ \citenamefont
  {Zhang}}]{wang2011efficient}%
  \BibitemOpen
  \bibfield  {author} {\bibinfo {author} {\bibfnamefont {W.-M.}\ \bibnamefont
  {Wang}}, \bibinfo {author} {\bibfnamefont {S.}~\bibnamefont {Kawata}},
  \bibinfo {author} {\bibfnamefont {Z.-M.}\ \bibnamefont {Sheng}}, \bibinfo
  {author} {\bibfnamefont {Y.-T.}\ \bibnamefont {Li}}, \bibinfo {author}
  {\bibfnamefont {L.-M.}\ \bibnamefont {Chen}}, \bibinfo {author}
  {\bibfnamefont {L.-J.}\ \bibnamefont {Qian}}, \ and\ \bibinfo {author}
  {\bibfnamefont {J.}~\bibnamefont {Zhang}},\ }\href {\doibase
  10.1364/OL.36.002608} {\bibfield  {journal} {\bibinfo  {journal} {Optics
  letters}\ }\textbf {\bibinfo {volume} {36}},\ \bibinfo {pages} {2608}
  (\bibinfo {year} {2011})}\BibitemShut {NoStop}%
\bibitem [{\citenamefont {D{\'e}chard}\ \emph {et~al.}(2019)\citenamefont
  {D{\'e}chard}, \citenamefont {Davoine},\ and\ \citenamefont
  {Berg{\'e}}}]{dechard2019thz}%
  \BibitemOpen
  \bibfield  {author} {\bibinfo {author} {\bibfnamefont {J.}~\bibnamefont
  {D{\'e}chard}}, \bibinfo {author} {\bibfnamefont {X.}~\bibnamefont
  {Davoine}}, \ and\ \bibinfo {author} {\bibfnamefont {L.}~\bibnamefont
  {Berg{\'e}}},\ }\href {\doibase 10.1103/PhysRevLett.123.264801} {\bibfield
  {journal} {\bibinfo  {journal} {Physical Review Letters}\ }\textbf {\bibinfo
  {volume} {123}},\ \bibinfo {pages} {264801} (\bibinfo {year}
  {2019})}\BibitemShut {NoStop}%
\bibitem [{\citenamefont {Andreev}\ \emph {et~al.}(1992)\citenamefont
  {Andreev}, \citenamefont {Gorbunov}, \citenamefont {Kirsanov}, \citenamefont
  {Pogosova},\ and\ \citenamefont {Ramazashvili}}]{andreev1992resonant}%
  \BibitemOpen
  \bibfield  {author} {\bibinfo {author} {\bibfnamefont {N.}~\bibnamefont
  {Andreev}}, \bibinfo {author} {\bibfnamefont {L.}~\bibnamefont {Gorbunov}},
  \bibinfo {author} {\bibfnamefont {V.}~\bibnamefont {Kirsanov}}, \bibinfo
  {author} {\bibfnamefont {A.}~\bibnamefont {Pogosova}}, \ and\ \bibinfo
  {author} {\bibfnamefont {R.}~\bibnamefont {Ramazashvili}},\ }\href@noop {}
  {\bibfield  {journal} {\bibinfo  {journal} {JETP lett}\ }\textbf {\bibinfo
  {volume} {55}},\ \bibinfo {pages} {571} (\bibinfo {year} {1992})}\BibitemShut
  {NoStop}%
\bibitem [{\citenamefont {Krall}\ \emph {et~al.}(1993)\citenamefont {Krall},
  \citenamefont {Ting}, \citenamefont {Esarey},\ and\ \citenamefont
  {Sprangle}}]{krall1993enhanced}%
  \BibitemOpen
  \bibfield  {author} {\bibinfo {author} {\bibfnamefont {J.}~\bibnamefont
  {Krall}}, \bibinfo {author} {\bibfnamefont {A.}~\bibnamefont {Ting}},
  \bibinfo {author} {\bibfnamefont {E.}~\bibnamefont {Esarey}}, \ and\ \bibinfo
  {author} {\bibfnamefont {P.}~\bibnamefont {Sprangle}},\ }\href@noop {}
  {\bibfield  {journal} {\bibinfo  {journal} {Physical Review E}\ }\textbf
  {\bibinfo {volume} {48}},\ \bibinfo {pages} {2157} (\bibinfo {year}
  {1993})}\BibitemShut {NoStop}%
\bibitem [{\citenamefont {Esarey}\ \emph {et~al.}(1994)\citenamefont {Esarey},
  \citenamefont {Krall},\ and\ \citenamefont {Sprangle}}]{PhysRevLett.72.2887}%
  \BibitemOpen
  \bibfield  {author} {\bibinfo {author} {\bibfnamefont {E.}~\bibnamefont
  {Esarey}}, \bibinfo {author} {\bibfnamefont {J.}~\bibnamefont {Krall}}, \
  and\ \bibinfo {author} {\bibfnamefont {P.}~\bibnamefont {Sprangle}},\ }\href
  {\doibase 10.1103/PhysRevLett.72.2887} {\bibfield  {journal} {\bibinfo
  {journal} {Phys. Rev. Lett.}\ }\textbf {\bibinfo {volume} {72}},\ \bibinfo
  {pages} {2887} (\bibinfo {year} {1994})}\BibitemShut {NoStop}%
\bibitem [{\citenamefont {Bulanov}\ \emph {et~al.}(1995)\citenamefont
  {Bulanov}, \citenamefont {Pegoraro},\ and\ \citenamefont
  {Pukhov}}]{bulanov1995two}%
  \BibitemOpen
  \bibfield  {author} {\bibinfo {author} {\bibfnamefont {S.}~\bibnamefont
  {Bulanov}}, \bibinfo {author} {\bibfnamefont {F.}~\bibnamefont {Pegoraro}}, \
  and\ \bibinfo {author} {\bibfnamefont {A.}~\bibnamefont {Pukhov}},\
  }\href@noop {} {\bibfield  {journal} {\bibinfo  {journal} {Physical review
  letters}\ }\textbf {\bibinfo {volume} {74}},\ \bibinfo {pages} {710}
  (\bibinfo {year} {1995})}\BibitemShut {NoStop}%
\bibitem [{\citenamefont {Le~Blanc}\ \emph {et~al.}(1997)\citenamefont
  {Le~Blanc}, \citenamefont {Downer}, \citenamefont {Wagner}, \citenamefont
  {Chen}, \citenamefont {Maksimchuk}, \citenamefont {Mourou},\ and\
  \citenamefont {Umstadter}}]{le1997temporal}%
  \BibitemOpen
  \bibfield  {author} {\bibinfo {author} {\bibfnamefont {S.}~\bibnamefont
  {Le~Blanc}}, \bibinfo {author} {\bibfnamefont {M.}~\bibnamefont {Downer}},
  \bibinfo {author} {\bibfnamefont {R.}~\bibnamefont {Wagner}}, \bibinfo
  {author} {\bibfnamefont {S.-Y.}\ \bibnamefont {Chen}}, \bibinfo {author}
  {\bibfnamefont {A.}~\bibnamefont {Maksimchuk}}, \bibinfo {author}
  {\bibfnamefont {G.}~\bibnamefont {Mourou}}, \ and\ \bibinfo {author}
  {\bibfnamefont {D.}~\bibnamefont {Umstadter}},\ }in\ \href@noop {} {\emph
  {\bibinfo {booktitle} {AIP Conference Proceedings}}},\ Vol.\ \bibinfo
  {volume} {398}\ (\bibinfo {organization} {American Institute of Physics},\
  \bibinfo {year} {1997})\ pp.\ \bibinfo {pages} {651--663}\BibitemShut
  {NoStop}%
\bibitem [{\citenamefont {Polyanskiy}\ \emph {et~al.}(2015)\citenamefont
  {Polyanskiy}, \citenamefont {Babzien},\ and\ \citenamefont
  {Pogorelsky}}]{polyanskiy2015chirped}%
  \BibitemOpen
  \bibfield  {author} {\bibinfo {author} {\bibfnamefont {M.~N.}\ \bibnamefont
  {Polyanskiy}}, \bibinfo {author} {\bibfnamefont {M.}~\bibnamefont {Babzien}},
  \ and\ \bibinfo {author} {\bibfnamefont {I.~V.}\ \bibnamefont {Pogorelsky}},\
  }\href {\doibase 10.1364/OPTICA.2.000675} {\bibfield  {journal} {\bibinfo
  {journal} {Optica}\ }\textbf {\bibinfo {volume} {2}},\ \bibinfo {pages} {675}
  (\bibinfo {year} {2015})}\BibitemShut {NoStop}%
\bibitem [{\citenamefont {Polyanskiy}\ \emph {et~al.}(2020)\citenamefont
  {Polyanskiy}, \citenamefont {Pogorelsky}, \citenamefont {Babzien},\ and\
  \citenamefont {Palmer}}]{polyanskiy2020demonstration}%
  \BibitemOpen
  \bibfield  {author} {\bibinfo {author} {\bibfnamefont {M.~N.}\ \bibnamefont
  {Polyanskiy}}, \bibinfo {author} {\bibfnamefont {I.~V.}\ \bibnamefont
  {Pogorelsky}}, \bibinfo {author} {\bibfnamefont {M.}~\bibnamefont {Babzien}},
  \ and\ \bibinfo {author} {\bibfnamefont {M.~A.}\ \bibnamefont {Palmer}},\
  }\href {\doibase 10.1364/OSAC.381467} {\bibfield  {journal} {\bibinfo
  {journal} {OSA Continuum}\ }\textbf {\bibinfo {volume} {3}},\ \bibinfo
  {pages} {459} (\bibinfo {year} {2020})}\BibitemShut {NoStop}%
\bibitem [{\citenamefont {Panagiotopoulos}\ \emph {et~al.}(2020)\citenamefont
  {Panagiotopoulos}, \citenamefont {Hastings}, \citenamefont {Kolesik},
  \citenamefont {Tochitsky},\ and\ \citenamefont
  {Moloney}}]{panagiotopoulos2020multi}%
  \BibitemOpen
  \bibfield  {author} {\bibinfo {author} {\bibfnamefont {P.}~\bibnamefont
  {Panagiotopoulos}}, \bibinfo {author} {\bibfnamefont {M.~G.}\ \bibnamefont
  {Hastings}}, \bibinfo {author} {\bibfnamefont {M.}~\bibnamefont {Kolesik}},
  \bibinfo {author} {\bibfnamefont {S.}~\bibnamefont {Tochitsky}}, \ and\
  \bibinfo {author} {\bibfnamefont {J.~V.}\ \bibnamefont {Moloney}},\ }\href
  {\doibase 10.1364/OSAC.399992} {\bibfield  {journal} {\bibinfo  {journal}
  {OSA Continuum}\ }\textbf {\bibinfo {volume} {3}},\ \bibinfo {pages} {3040}
  (\bibinfo {year} {2020})}\BibitemShut {NoStop}%
\bibitem [{\citenamefont {Kumar}\ \emph {et~al.}(2019)\citenamefont {Kumar},
  \citenamefont {Yu}, \citenamefont {Zgadzaj}, \citenamefont {Amorim},
  \citenamefont {Downer}, \citenamefont {Welch}, \citenamefont {Litvinenko},
  \citenamefont {Vafaei-Najafabadi},\ and\ \citenamefont
  {Samulyak}}]{kumar2019simulation}%
  \BibitemOpen
  \bibfield  {author} {\bibinfo {author} {\bibfnamefont {P.}~\bibnamefont
  {Kumar}}, \bibinfo {author} {\bibfnamefont {K.}~\bibnamefont {Yu}}, \bibinfo
  {author} {\bibfnamefont {R.}~\bibnamefont {Zgadzaj}}, \bibinfo {author}
  {\bibfnamefont {L.~D.}\ \bibnamefont {Amorim}}, \bibinfo {author}
  {\bibfnamefont {M.}~\bibnamefont {Downer}}, \bibinfo {author} {\bibfnamefont
  {J.}~\bibnamefont {Welch}}, \bibinfo {author} {\bibfnamefont {V.~N.}\
  \bibnamefont {Litvinenko}}, \bibinfo {author} {\bibfnamefont
  {N.}~\bibnamefont {Vafaei-Najafabadi}}, \ and\ \bibinfo {author}
  {\bibfnamefont {R.}~\bibnamefont {Samulyak}},\ }\href {\doibase
  10.1063/1.5095780} {\bibfield  {journal} {\bibinfo  {journal} {Physics of
  Plasmas}\ }\textbf {\bibinfo {volume} {26}} (\bibinfo {year} {2019}),\
  10.1063/1.5095780}\BibitemShut {NoStop}%
\bibitem [{\citenamefont {Brunetti}\ \emph {et~al.}(2022)\citenamefont
  {Brunetti}, \citenamefont {Campbell}, \citenamefont {Lovell},\ and\
  \citenamefont {Jaroszynski}}]{brunetti2022high}%
  \BibitemOpen
  \bibfield  {author} {\bibinfo {author} {\bibfnamefont {E.}~\bibnamefont
  {Brunetti}}, \bibinfo {author} {\bibfnamefont {R.~N.}\ \bibnamefont
  {Campbell}}, \bibinfo {author} {\bibfnamefont {J.}~\bibnamefont {Lovell}}, \
  and\ \bibinfo {author} {\bibfnamefont {D.~A.}\ \bibnamefont {Jaroszynski}},\
  }\href@noop {} {\bibfield  {journal} {\bibinfo  {journal} {Scientific
  Reports}\ }\textbf {\bibinfo {volume} {12}},\ \bibinfo {pages} {6703}
  (\bibinfo {year} {2022})}\BibitemShut {NoStop}%
\bibitem [{\citenamefont {Pogorelsky}\ \emph
  {et~al.}(2016{\natexlab{a}})\citenamefont {Pogorelsky}, \citenamefont
  {Polyanskiy},\ and\ \citenamefont {Kimura}}]{pogorelsky2016mid}%
  \BibitemOpen
  \bibfield  {author} {\bibinfo {author} {\bibfnamefont {I.}~\bibnamefont
  {Pogorelsky}}, \bibinfo {author} {\bibfnamefont {M.}~\bibnamefont
  {Polyanskiy}}, \ and\ \bibinfo {author} {\bibfnamefont {W.}~\bibnamefont
  {Kimura}},\ }\href@noop {} {\bibfield  {journal} {\bibinfo  {journal}
  {Physical Review Accelerators and Beams}\ }\textbf {\bibinfo {volume} {19}},\
  \bibinfo {pages} {091001} (\bibinfo {year} {2016}{\natexlab{a}})}\BibitemShut
  {NoStop}%
\bibitem [{\citenamefont {Pogorelsky}\ \emph
  {et~al.}(2016{\natexlab{b}})\citenamefont {Pogorelsky}, \citenamefont
  {Babzien}, \citenamefont {Ben-Zvi}, \citenamefont {Skaritka},\ and\
  \citenamefont {Polyanskiy}}]{pogorelsky2016bestia}%
  \BibitemOpen
  \bibfield  {author} {\bibinfo {author} {\bibfnamefont {I.~V.}\ \bibnamefont
  {Pogorelsky}}, \bibinfo {author} {\bibfnamefont {M.}~\bibnamefont {Babzien}},
  \bibinfo {author} {\bibfnamefont {I.}~\bibnamefont {Ben-Zvi}}, \bibinfo
  {author} {\bibfnamefont {J.}~\bibnamefont {Skaritka}}, \ and\ \bibinfo
  {author} {\bibfnamefont {M.~N.}\ \bibnamefont {Polyanskiy}},\ }\href@noop {}
  {\bibfield  {journal} {\bibinfo  {journal} {Nuclear Instruments and Methods
  in Physics Research Section A: Accelerators, Spectrometers, Detectors and
  Associated Equipment}\ }\textbf {\bibinfo {volume} {829}},\ \bibinfo {pages}
  {432} (\bibinfo {year} {2016}{\natexlab{b}})}\BibitemShut {NoStop}%
\bibitem [{\citenamefont {Maity}\ \emph {et~al.}(2021)\citenamefont {Maity},
  \citenamefont {Mandal}, \citenamefont {Vashistha}, \citenamefont {Goswami},\
  and\ \citenamefont {Das}}]{maity2021harmonic}%
  \BibitemOpen
  \bibfield  {author} {\bibinfo {author} {\bibfnamefont {S.}~\bibnamefont
  {Maity}}, \bibinfo {author} {\bibfnamefont {D.}~\bibnamefont {Mandal}},
  \bibinfo {author} {\bibfnamefont {A.}~\bibnamefont {Vashistha}}, \bibinfo
  {author} {\bibfnamefont {L.~P.}\ \bibnamefont {Goswami}}, \ and\ \bibinfo
  {author} {\bibfnamefont {A.}~\bibnamefont {Das}},\ }\href@noop {} {\bibfield
  {journal} {\bibinfo  {journal} {Journal of Plasma Physics}\ }\textbf
  {\bibinfo {volume} {87}},\ \bibinfo {pages} {905870509} (\bibinfo {year}
  {2021})}\BibitemShut {NoStop}%
\bibitem [{\citenamefont {Maity}\ \emph {et~al.}(2022)\citenamefont {Maity},
  \citenamefont {Goswami}, \citenamefont {Vashistha}, \citenamefont {Mandal},\
  and\ \citenamefont {Das}}]{maity2022mode}%
  \BibitemOpen
  \bibfield  {author} {\bibinfo {author} {\bibfnamefont {S.}~\bibnamefont
  {Maity}}, \bibinfo {author} {\bibfnamefont {L.~P.}\ \bibnamefont {Goswami}},
  \bibinfo {author} {\bibfnamefont {A.}~\bibnamefont {Vashistha}}, \bibinfo
  {author} {\bibfnamefont {D.}~\bibnamefont {Mandal}}, \ and\ \bibinfo {author}
  {\bibfnamefont {A.}~\bibnamefont {Das}},\ }\href@noop {} {\bibfield
  {journal} {\bibinfo  {journal} {Physical Review E}\ }\textbf {\bibinfo
  {volume} {105}},\ \bibinfo {pages} {055209} (\bibinfo {year}
  {2022})}\BibitemShut {NoStop}%
\bibitem [{\citenamefont {Juneja}\ \emph {et~al.}(2023)\citenamefont {Juneja},
  \citenamefont {Dhalia}, \citenamefont {Goswami}, \citenamefont {Maity},
  \citenamefont {Mandal},\ and\ \citenamefont {Das}}]{juneja2023ion}%
  \BibitemOpen
  \bibfield  {author} {\bibinfo {author} {\bibfnamefont {R.}~\bibnamefont
  {Juneja}}, \bibinfo {author} {\bibfnamefont {T.}~\bibnamefont {Dhalia}},
  \bibinfo {author} {\bibfnamefont {L.~P.}\ \bibnamefont {Goswami}}, \bibinfo
  {author} {\bibfnamefont {S.}~\bibnamefont {Maity}}, \bibinfo {author}
  {\bibfnamefont {D.}~\bibnamefont {Mandal}}, \ and\ \bibinfo {author}
  {\bibfnamefont {A.}~\bibnamefont {Das}},\ }\href@noop {} {\bibfield
  {journal} {\bibinfo  {journal} {Plasma Physics and Controlled Fusion}\
  }\textbf {\bibinfo {volume} {65}},\ \bibinfo {pages} {095005} (\bibinfo
  {year} {2023})}\BibitemShut {NoStop}%
\bibitem [{\citenamefont {Dhalia}\ \emph {et~al.}(2023)\citenamefont {Dhalia},
  \citenamefont {Juneja}, \citenamefont {Goswami}, \citenamefont {Maity},\ and\
  \citenamefont {Das}}]{dhalia2023harmonic}%
  \BibitemOpen
  \bibfield  {author} {\bibinfo {author} {\bibfnamefont {T.}~\bibnamefont
  {Dhalia}}, \bibinfo {author} {\bibfnamefont {R.}~\bibnamefont {Juneja}},
  \bibinfo {author} {\bibfnamefont {L.~P.}\ \bibnamefont {Goswami}}, \bibinfo
  {author} {\bibfnamefont {S.}~\bibnamefont {Maity}}, \ and\ \bibinfo {author}
  {\bibfnamefont {A.}~\bibnamefont {Das}},\ }\href@noop {} {\bibfield
  {journal} {\bibinfo  {journal} {Journal of Physics D: Applied Physics}\
  }\textbf {\bibinfo {volume} {56}},\ \bibinfo {pages} {395201} (\bibinfo
  {year} {2023})}\BibitemShut {NoStop}%
\bibitem [{\citenamefont {Vashistha}\ \emph {et~al.}(2023)\citenamefont
  {Vashistha}, \citenamefont {Mandal}, \citenamefont {Maity},\ and\
  \citenamefont {Das}}]{vashistha2023localized}%
  \BibitemOpen
  \bibfield  {author} {\bibinfo {author} {\bibfnamefont {A.}~\bibnamefont
  {Vashistha}}, \bibinfo {author} {\bibfnamefont {D.}~\bibnamefont {Mandal}},
  \bibinfo {author} {\bibfnamefont {S.}~\bibnamefont {Maity}}, \ and\ \bibinfo
  {author} {\bibfnamefont {A.}~\bibnamefont {Das}},\ }\href@noop {} {\bibfield
  {journal} {\bibinfo  {journal} {Plasma Physics and Controlled Fusion}\
  }\textbf {\bibinfo {volume} {65}},\ \bibinfo {pages} {035006} (\bibinfo
  {year} {2023})}\BibitemShut {NoStop}%
\bibitem [{\citenamefont {Estabrook}\ \emph {et~al.}(1980)\citenamefont
  {Estabrook}, \citenamefont {Kruer},\ and\ \citenamefont
  {Lasinski}}]{estabrook1980heating}%
  \BibitemOpen
  \bibfield  {author} {\bibinfo {author} {\bibfnamefont {K.}~\bibnamefont
  {Estabrook}}, \bibinfo {author} {\bibfnamefont {W.}~\bibnamefont {Kruer}}, \
  and\ \bibinfo {author} {\bibfnamefont {B.}~\bibnamefont {Lasinski}},\
  }\href@noop {} {\bibfield  {journal} {\bibinfo  {journal} {Physical Review
  Letters}\ }\textbf {\bibinfo {volume} {45}},\ \bibinfo {pages} {1399}
  (\bibinfo {year} {1980})}\BibitemShut {NoStop}%
\bibitem [{\citenamefont {Joshi}\ \emph {et~al.}(1981)\citenamefont {Joshi},
  \citenamefont {Tajima}, \citenamefont {Dawson}, \citenamefont {Baldis},\ and\
  \citenamefont {Ebrahim}}]{PhysRevLett.47.1285}%
  \BibitemOpen
  \bibfield  {author} {\bibinfo {author} {\bibfnamefont {C.}~\bibnamefont
  {Joshi}}, \bibinfo {author} {\bibfnamefont {T.}~\bibnamefont {Tajima}},
  \bibinfo {author} {\bibfnamefont {J.~M.}\ \bibnamefont {Dawson}}, \bibinfo
  {author} {\bibfnamefont {H.~A.}\ \bibnamefont {Baldis}}, \ and\ \bibinfo
  {author} {\bibfnamefont {N.~A.}\ \bibnamefont {Ebrahim}},\ }\href {\doibase
  10.1103/PhysRevLett.47.1285} {\bibfield  {journal} {\bibinfo  {journal}
  {Phys. Rev. Lett.}\ }\textbf {\bibinfo {volume} {47}},\ \bibinfo {pages}
  {1285} (\bibinfo {year} {1981})}\BibitemShut {NoStop}%
\bibitem [{\citenamefont {Mori}\ \emph {et~al.}(1994)\citenamefont {Mori},
  \citenamefont {Decker}, \citenamefont {Hinkel},\ and\ \citenamefont
  {Katsouleas}}]{PhysRevLett.72.1482}%
  \BibitemOpen
  \bibfield  {author} {\bibinfo {author} {\bibfnamefont {W.~B.}\ \bibnamefont
  {Mori}}, \bibinfo {author} {\bibfnamefont {C.~D.}\ \bibnamefont {Decker}},
  \bibinfo {author} {\bibfnamefont {D.~E.}\ \bibnamefont {Hinkel}}, \ and\
  \bibinfo {author} {\bibfnamefont {T.}~\bibnamefont {Katsouleas}},\ }\href
  {\doibase 10.1103/PhysRevLett.72.1482} {\bibfield  {journal} {\bibinfo
  {journal} {Phys. Rev. Lett.}\ }\textbf {\bibinfo {volume} {72}},\ \bibinfo
  {pages} {1482} (\bibinfo {year} {1994})}\BibitemShut {NoStop}%
\bibitem [{\citenamefont {Fisher}\ and\ \citenamefont
  {Tajima}(1996)}]{fisher1996enhanced}%
  \BibitemOpen
  \bibfield  {author} {\bibinfo {author} {\bibfnamefont {D.}~\bibnamefont
  {Fisher}}\ and\ \bibinfo {author} {\bibfnamefont {T.}~\bibnamefont
  {Tajima}},\ }\href@noop {} {\bibfield  {journal} {\bibinfo  {journal}
  {Physical Review E}\ }\textbf {\bibinfo {volume} {53}},\ \bibinfo {pages}
  {1844} (\bibinfo {year} {1996})}\BibitemShut {NoStop}%
\bibitem [{\citenamefont {Najmudin}\ \emph {et~al.}(2000)\citenamefont
  {Najmudin}, \citenamefont {Allott}, \citenamefont {Amiranoff}, \citenamefont
  {Clark}, \citenamefont {Danson}, \citenamefont {Gordon}, \citenamefont
  {Joshi}, \citenamefont {Krushelnick}, \citenamefont {Malka}, \citenamefont
  {Neely} \emph {et~al.}}]{najmudin2000measurement}%
  \BibitemOpen
  \bibfield  {author} {\bibinfo {author} {\bibfnamefont {Z.}~\bibnamefont
  {Najmudin}}, \bibinfo {author} {\bibfnamefont {R.}~\bibnamefont {Allott}},
  \bibinfo {author} {\bibfnamefont {F.}~\bibnamefont {Amiranoff}}, \bibinfo
  {author} {\bibfnamefont {E.}~\bibnamefont {Clark}}, \bibinfo {author}
  {\bibfnamefont {C.}~\bibnamefont {Danson}}, \bibinfo {author} {\bibfnamefont
  {D.~F.}\ \bibnamefont {Gordon}}, \bibinfo {author} {\bibfnamefont
  {C.}~\bibnamefont {Joshi}}, \bibinfo {author} {\bibfnamefont
  {K.}~\bibnamefont {Krushelnick}}, \bibinfo {author} {\bibfnamefont
  {V.}~\bibnamefont {Malka}}, \bibinfo {author} {\bibfnamefont
  {D.}~\bibnamefont {Neely}},  \emph {et~al.},\ }\href@noop {} {\bibfield
  {journal} {\bibinfo  {journal} {IEEE transactions on plasma science}\
  }\textbf {\bibinfo {volume} {28}},\ \bibinfo {pages} {1122} (\bibinfo {year}
  {2000})}\BibitemShut {NoStop}%
\bibitem [{\citenamefont {Arber}\ \emph {et~al.}(2015)\citenamefont {Arber},
  \citenamefont {Bennett}, \citenamefont {Brady}, \citenamefont
  {Lawrence-Douglas}, \citenamefont {Ramsay}, \citenamefont {Sircombe},
  \citenamefont {Gillies}, \citenamefont {Evans}, \citenamefont {Schmitz},
  \citenamefont {Bell} \emph {et~al.}}]{arber2015contemporary}%
  \BibitemOpen
  \bibfield  {author} {\bibinfo {author} {\bibfnamefont {T.~D.}\ \bibnamefont
  {Arber}}, \bibinfo {author} {\bibfnamefont {K.}~\bibnamefont {Bennett}},
  \bibinfo {author} {\bibfnamefont {C.~S.}\ \bibnamefont {Brady}}, \bibinfo
  {author} {\bibfnamefont {A.}~\bibnamefont {Lawrence-Douglas}}, \bibinfo
  {author} {\bibfnamefont {M.~G.}\ \bibnamefont {Ramsay}}, \bibinfo {author}
  {\bibfnamefont {N.~J.}\ \bibnamefont {Sircombe}}, \bibinfo {author}
  {\bibfnamefont {P.}~\bibnamefont {Gillies}}, \bibinfo {author} {\bibfnamefont
  {R.~G.}\ \bibnamefont {Evans}}, \bibinfo {author} {\bibfnamefont
  {H.}~\bibnamefont {Schmitz}}, \bibinfo {author} {\bibfnamefont {A.~R.}\
  \bibnamefont {Bell}},  \emph {et~al.},\ }\href {\doibase
  10.1088/0741-3335/57/11/113001} {\bibfield  {journal} {\bibinfo  {journal}
  {Plasma Physics and Controlled Fusion}\ }\textbf {\bibinfo {volume} {57}},\
  \bibinfo {pages} {113001} (\bibinfo {year} {2015})}\BibitemShut {NoStop}%
\bibitem [{\citenamefont {Bennett}\ \emph {et~al.}(2017)\citenamefont
  {Bennett}, \citenamefont {Brady}, \citenamefont {Schmitz}, \citenamefont
  {Ridgers}, \citenamefont {Arber}, \citenamefont {Evans},\ and\ \citenamefont
  {Bell}}]{bennett2017users}%
  \BibitemOpen
  \bibfield  {author} {\bibinfo {author} {\bibfnamefont {K.}~\bibnamefont
  {Bennett}}, \bibinfo {author} {\bibfnamefont {C.}~\bibnamefont {Brady}},
  \bibinfo {author} {\bibfnamefont {H.}~\bibnamefont {Schmitz}}, \bibinfo
  {author} {\bibfnamefont {C.}~\bibnamefont {Ridgers}}, \bibinfo {author}
  {\bibfnamefont {T.}~\bibnamefont {Arber}}, \bibinfo {author} {\bibfnamefont
  {R.}~\bibnamefont {Evans}}, \ and\ \bibinfo {author} {\bibfnamefont
  {T.}~\bibnamefont {Bell}},\ }\href
  {https://www.archie-west.ac.uk/wp-content/uploads/2014/02/epoch_user-4.3.pdf}
  {\bibfield  {journal} {\bibinfo  {journal} {University of Warwick}\ }
  (\bibinfo {year} {2017})}\BibitemShut {NoStop}%
\bibitem [{\citenamefont {Yee}(1966)}]{yee1966numerical}%
  \BibitemOpen
  \bibfield  {author} {\bibinfo {author} {\bibfnamefont {K.}~\bibnamefont
  {Yee}},\ }\href {\doibase 10.1109/TAP.1966.1138693} {\bibfield  {journal}
  {\bibinfo  {journal} {IEEE Transactions on antennas and propagation}\
  }\textbf {\bibinfo {volume} {14}},\ \bibinfo {pages} {302} (\bibinfo {year}
  {1966})}\BibitemShut {NoStop}%
\bibitem [{\citenamefont {Boris}\ \emph {et~al.}(1970)\citenamefont {Boris}
  \emph {et~al.}}]{boris1970relativistic}%
  \BibitemOpen
  \bibfield  {author} {\bibinfo {author} {\bibfnamefont {J.~P.}\ \bibnamefont
  {Boris}} \emph {et~al.},\ }in\ \href@noop {} {\emph {\bibinfo {booktitle}
  {Proc. 4th Conf. on Numerical Simulation of Plasmas (Washington, DC)}}}\
  (\bibinfo {year} {1970})\ pp.\ \bibinfo {pages} {3--67}\BibitemShut {NoStop}%
\bibitem [{\citenamefont {Gibbon}(2005)}]{gibbon2005short}%
  \BibitemOpen
  \bibfield  {author} {\bibinfo {author} {\bibfnamefont {P.}~\bibnamefont
  {Gibbon}},\ }\href {\doibase 10.1142/p116} {\emph {\bibinfo {title} {Short
  pulse laser interactions with matter: an introduction}}}\ (\bibinfo
  {publisher} {World Scientific},\ \bibinfo {year} {2005})\BibitemShut
  {NoStop}%
\bibitem [{\citenamefont {Delone}\ and\ \citenamefont
  {Krainov}(2000)}]{delone2000multiphoton}%
  \BibitemOpen
  \bibfield  {author} {\bibinfo {author} {\bibfnamefont {N.~B.}\ \bibnamefont
  {Delone}}\ and\ \bibinfo {author} {\bibfnamefont {V.~P.}\ \bibnamefont
  {Krainov}},\ }\href {\doibase 10.1007/978-3-642-57208-1} {\emph {\bibinfo
  {title} {Multiphoton processes in atoms}}},\ Vol.~\bibinfo {volume} {13}\
  (\bibinfo  {publisher} {Springer Berlin, Heidelberg},\ \bibinfo {year}
  {2000})\BibitemShut {NoStop}%
\bibitem [{\citenamefont {Ammosov}\ \emph {et~al.}(1986)\citenamefont
  {Ammosov}, \citenamefont {Delone},\ and\ \citenamefont
  {Krainov}}]{ammosov1986tunnel}%
  \BibitemOpen
  \bibfield  {author} {\bibinfo {author} {\bibfnamefont {M.~V.}\ \bibnamefont
  {Ammosov}}, \bibinfo {author} {\bibfnamefont {N.~B.}\ \bibnamefont {Delone}},
  \ and\ \bibinfo {author} {\bibfnamefont {V.~P.}\ \bibnamefont {Krainov}},\
  }\href {http://jetp.ras.ru/cgi-bin/dn/e_064_06_1191} {\bibfield  {journal}
  {\bibinfo  {journal} {Soviet Journal of Experimental and Theoretical
  Physics}\ }\textbf {\bibinfo {volume} {64}},\ \bibinfo {pages} {1191}
  (\bibinfo {year} {1986})}\BibitemShut {NoStop}%
\bibitem [{\citenamefont {Krainov}(1995)}]{krainov1995theory}%
  \BibitemOpen
  \bibfield  {author} {\bibinfo {author} {\bibfnamefont {V.~P.}\ \bibnamefont
  {Krainov}},\ }\href {\doibase 10.1142/S0218863595000343} {\bibfield
  {journal} {\bibinfo  {journal} {Journal of Nonlinear Optical Physics \&
  Materials}\ }\textbf {\bibinfo {volume} {4}},\ \bibinfo {pages} {775}
  (\bibinfo {year} {1995})}\BibitemShut {NoStop}%
\bibitem [{\citenamefont {Kruer}(1988)}]{kruer1988physics}%
  \BibitemOpen
  \bibfield  {author} {\bibinfo {author} {\bibfnamefont {W.~L.}\ \bibnamefont
  {Kruer}},\ }\href {https://worldcat.org/title/16276733} {\emph {\bibinfo
  {title} {The Physics Of Laser Plasma Interactions}}}\ (\bibinfo  {publisher}
  {Addison-Wesley, Redwood City, Calif.},\ \bibinfo {year} {1988})\BibitemShut
  {NoStop}%
\bibitem [{\citenamefont {Mori}\ and\ \citenamefont
  {Katsouleas}(1992)}]{mori1992ponderomotive}%
  \BibitemOpen
  \bibfield  {author} {\bibinfo {author} {\bibfnamefont {W.}~\bibnamefont
  {Mori}}\ and\ \bibinfo {author} {\bibfnamefont {T.}~\bibnamefont
  {Katsouleas}},\ }\href {\doibase 10.1103/PhysRevLett.69.3495} {\bibfield
  {journal} {\bibinfo  {journal} {Physical review letters}\ }\textbf {\bibinfo
  {volume} {69}},\ \bibinfo {pages} {3495} (\bibinfo {year}
  {1992})}\BibitemShut {NoStop}%
\bibitem [{\citenamefont {Albert}\ \emph {et~al.}(2017)\citenamefont {Albert},
  \citenamefont {Lemos}, \citenamefont {Shaw}, \citenamefont {Pollock},
  \citenamefont {Goyon}, \citenamefont {Schumaker}, \citenamefont {Saunders},
  \citenamefont {Marsh}, \citenamefont {Pak}, \citenamefont {Ralph},
  \citenamefont {Martins}, \citenamefont {Amorim}, \citenamefont {Falcone},
  \citenamefont {Glenzer}, \citenamefont {Moody},\ and\ \citenamefont
  {Joshi}}]{PhysRevLett.118.134801}%
  \BibitemOpen
  \bibfield  {author} {\bibinfo {author} {\bibfnamefont {F.}~\bibnamefont
  {Albert}}, \bibinfo {author} {\bibfnamefont {N.}~\bibnamefont {Lemos}},
  \bibinfo {author} {\bibfnamefont {J.~L.}\ \bibnamefont {Shaw}}, \bibinfo
  {author} {\bibfnamefont {B.~B.}\ \bibnamefont {Pollock}}, \bibinfo {author}
  {\bibfnamefont {C.}~\bibnamefont {Goyon}}, \bibinfo {author} {\bibfnamefont
  {W.}~\bibnamefont {Schumaker}}, \bibinfo {author} {\bibfnamefont {A.~M.}\
  \bibnamefont {Saunders}}, \bibinfo {author} {\bibfnamefont {K.~A.}\
  \bibnamefont {Marsh}}, \bibinfo {author} {\bibfnamefont {A.}~\bibnamefont
  {Pak}}, \bibinfo {author} {\bibfnamefont {J.~E.}\ \bibnamefont {Ralph}},
  \bibinfo {author} {\bibfnamefont {J.~L.}\ \bibnamefont {Martins}}, \bibinfo
  {author} {\bibfnamefont {L.~D.}\ \bibnamefont {Amorim}}, \bibinfo {author}
  {\bibfnamefont {R.~W.}\ \bibnamefont {Falcone}}, \bibinfo {author}
  {\bibfnamefont {S.~H.}\ \bibnamefont {Glenzer}}, \bibinfo {author}
  {\bibfnamefont {J.~D.}\ \bibnamefont {Moody}}, \ and\ \bibinfo {author}
  {\bibfnamefont {C.}~\bibnamefont {Joshi}},\ }\href {\doibase
  10.1103/PhysRevLett.118.134801} {\bibfield  {journal} {\bibinfo  {journal}
  {Phys. Rev. Lett.}\ }\textbf {\bibinfo {volume} {118}},\ \bibinfo {pages}
  {134801} (\bibinfo {year} {2017})}\BibitemShut {NoStop}%
\bibitem [{\citenamefont {Leemans}\ \emph {et~al.}(2003)\citenamefont
  {Leemans}, \citenamefont {Geddes}, \citenamefont {Faure}, \citenamefont
  {T\'oth}, \citenamefont {van Tilborg}, \citenamefont {Schroeder},
  \citenamefont {Esarey}, \citenamefont {Fubiani}, \citenamefont {Auerbach},
  \citenamefont {Marcelis}, \citenamefont {Carnahan}, \citenamefont {Kaindl},
  \citenamefont {Byrd},\ and\ \citenamefont {Martin}}]{PhysRevLett.91.074802}%
  \BibitemOpen
  \bibfield  {author} {\bibinfo {author} {\bibfnamefont {W.~P.}\ \bibnamefont
  {Leemans}}, \bibinfo {author} {\bibfnamefont {C.~G.~R.}\ \bibnamefont
  {Geddes}}, \bibinfo {author} {\bibfnamefont {J.}~\bibnamefont {Faure}},
  \bibinfo {author} {\bibfnamefont {C.}~\bibnamefont {T\'oth}}, \bibinfo
  {author} {\bibfnamefont {J.}~\bibnamefont {van Tilborg}}, \bibinfo {author}
  {\bibfnamefont {C.~B.}\ \bibnamefont {Schroeder}}, \bibinfo {author}
  {\bibfnamefont {E.}~\bibnamefont {Esarey}}, \bibinfo {author} {\bibfnamefont
  {G.}~\bibnamefont {Fubiani}}, \bibinfo {author} {\bibfnamefont
  {D.}~\bibnamefont {Auerbach}}, \bibinfo {author} {\bibfnamefont
  {B.}~\bibnamefont {Marcelis}}, \bibinfo {author} {\bibfnamefont {M.~A.}\
  \bibnamefont {Carnahan}}, \bibinfo {author} {\bibfnamefont {R.~A.}\
  \bibnamefont {Kaindl}}, \bibinfo {author} {\bibfnamefont {J.}~\bibnamefont
  {Byrd}}, \ and\ \bibinfo {author} {\bibfnamefont {M.~C.}\ \bibnamefont
  {Martin}},\ }\href {\doibase 10.1103/PhysRevLett.91.074802} {\bibfield
  {journal} {\bibinfo  {journal} {Phys. Rev. Lett.}\ }\textbf {\bibinfo
  {volume} {91}},\ \bibinfo {pages} {074802} (\bibinfo {year}
  {2003})}\BibitemShut {NoStop}%
\bibitem [{\citenamefont {Sprangle}\ \emph {et~al.}(1987)\citenamefont
  {Sprangle}, \citenamefont {Tang},\ and\ \citenamefont
  {Esarey}}]{sprangle1987relativistic}%
  \BibitemOpen
  \bibfield  {author} {\bibinfo {author} {\bibfnamefont {P.}~\bibnamefont
  {Sprangle}}, \bibinfo {author} {\bibfnamefont {C.-M.}\ \bibnamefont {Tang}},
  \ and\ \bibinfo {author} {\bibfnamefont {E.}~\bibnamefont {Esarey}},\ }\href
  {\doibase 10.1109/TPS.1987.4316677} {\bibfield  {journal} {\bibinfo
  {journal} {IEEE transactions on plasma science}\ }\textbf {\bibinfo {volume}
  {15}},\ \bibinfo {pages} {145} (\bibinfo {year} {1987})}\BibitemShut
  {NoStop}%
\bibitem [{\citenamefont {Sprangle}\ \emph {et~al.}(1990)\citenamefont
  {Sprangle}, \citenamefont {Esarey},\ and\ \citenamefont
  {Ting}}]{PhysRevA.41.4463}%
  \BibitemOpen
  \bibfield  {author} {\bibinfo {author} {\bibfnamefont {P.}~\bibnamefont
  {Sprangle}}, \bibinfo {author} {\bibfnamefont {E.}~\bibnamefont {Esarey}}, \
  and\ \bibinfo {author} {\bibfnamefont {A.}~\bibnamefont {Ting}},\ }\href
  {\doibase 10.1103/PhysRevA.41.4463} {\bibfield  {journal} {\bibinfo
  {journal} {Phys. Rev. A}\ }\textbf {\bibinfo {volume} {41}},\ \bibinfo
  {pages} {4463} (\bibinfo {year} {1990})}\BibitemShut {NoStop}%
\bibitem [{\citenamefont {Sun}\ \emph {et~al.}(1987)\citenamefont {Sun},
  \citenamefont {Ott}, \citenamefont {Lee},\ and\ \citenamefont
  {Guzdar}}]{sun1987self}%
  \BibitemOpen
  \bibfield  {author} {\bibinfo {author} {\bibfnamefont {G.-Z.}\ \bibnamefont
  {Sun}}, \bibinfo {author} {\bibfnamefont {E.}~\bibnamefont {Ott}}, \bibinfo
  {author} {\bibfnamefont {Y.}~\bibnamefont {Lee}}, \ and\ \bibinfo {author}
  {\bibfnamefont {P.}~\bibnamefont {Guzdar}},\ }\href {\doibase
  10.1063/1.866349} {\bibfield  {journal} {\bibinfo  {journal} {The Physics of
  fluids}\ }\textbf {\bibinfo {volume} {30}},\ \bibinfo {pages} {526} (\bibinfo
  {year} {1987})}\BibitemShut {NoStop}%
\bibitem [{\citenamefont {Kurki-Suonio}\ \emph {et~al.}(1989)\citenamefont
  {Kurki-Suonio}, \citenamefont {Morrison},\ and\ \citenamefont
  {Tajima}}]{PhysRevA.40.3230}%
  \BibitemOpen
  \bibfield  {author} {\bibinfo {author} {\bibfnamefont {T.}~\bibnamefont
  {Kurki-Suonio}}, \bibinfo {author} {\bibfnamefont {P.~J.}\ \bibnamefont
  {Morrison}}, \ and\ \bibinfo {author} {\bibfnamefont {T.}~\bibnamefont
  {Tajima}},\ }\href {\doibase 10.1103/PhysRevA.40.3230} {\bibfield  {journal}
  {\bibinfo  {journal} {Phys. Rev. A}\ }\textbf {\bibinfo {volume} {40}},\
  \bibinfo {pages} {3230} (\bibinfo {year} {1989})}\BibitemShut {NoStop}%
\bibitem [{\citenamefont {Lu}\ \emph {et~al.}(2007)\citenamefont {Lu},
  \citenamefont {Tzoufras}, \citenamefont {Joshi}, \citenamefont {Tsung},
  \citenamefont {Mori}, \citenamefont {Vieira}, \citenamefont {Fonseca},\ and\
  \citenamefont {Silva}}]{lu2007generating}%
  \BibitemOpen
  \bibfield  {author} {\bibinfo {author} {\bibfnamefont {W.}~\bibnamefont
  {Lu}}, \bibinfo {author} {\bibfnamefont {M.}~\bibnamefont {Tzoufras}},
  \bibinfo {author} {\bibfnamefont {C.}~\bibnamefont {Joshi}}, \bibinfo
  {author} {\bibfnamefont {F.}~\bibnamefont {Tsung}}, \bibinfo {author}
  {\bibfnamefont {W.}~\bibnamefont {Mori}}, \bibinfo {author} {\bibfnamefont
  {J.}~\bibnamefont {Vieira}}, \bibinfo {author} {\bibfnamefont
  {R.}~\bibnamefont {Fonseca}}, \ and\ \bibinfo {author} {\bibfnamefont
  {L.}~\bibnamefont {Silva}},\ }\href {\doibase 10.1103/PhysRevSTAB.10.061301}
  {\bibfield  {journal} {\bibinfo  {journal} {Physical Review Special
  Topics-Accelerators and Beams}\ }\textbf {\bibinfo {volume} {10}},\ \bibinfo
  {pages} {061301} (\bibinfo {year} {2007})}\BibitemShut {NoStop}%
\end{thebibliography}%
\end{document}